%% file: main.tex
\documentclass[%
 aip,
 amsmath,amssymb,
 reprint,%
]{revtex4-2}

\allowdisplaybreaks

\input{settings/operator_definitions}

\input{settings/setup_options}

\input{settings/tex_symbols}

\input{settings/user_defined}

\makeatletter
\def\@email#1#2{%
 \endgroup
 \patchcmd{\titleblock@produce}
  {\frontmatter@RRAPformat}
  {\frontmatter@RRAPformat{\produce@RRAP{*#1\href{mailto:#2}{#2}}}\frontmatter@RRAPformat}
  {}{}
}%
\makeatother
\begin{document}

\preprint{AIP/123-QED}

\title[RPA dielectric function approximation]{A new efficient and accurate approximation for the dielectric function of the weakly coupled electron gas}

\author{D.~A.~Chapman}
 \email{david.chapman@firstlightfusion.com.}
\affiliation{ 
First Light Fusion Ltd., Unit 9/10 Oxford Pioneer Park, Yarnton, OX5 1QU, United Kingdom
}

\date{\today}

\begin{abstract}
\input{sections/abstract}
\end{abstract}

\maketitle
\allowdisplaybreaks

\input{sections/main_body/01_introduction}

\input{sections/main_body/02_analytic_properties}

\input{sections/main_body/03_numerical_benchmark}

\input{sections/main_body/04_existing_approximations}
\input{sections/main_body/05_new_model}

\input{sections/main_body/06_results_and_applications}
\input{sections/main_body/07_conclusions}

\begin{appendix}
\input{sections/appendices/01_DAC_model_equations}

\input{sections/appendices/02_end_point_construction}

\input{sections/appendices/03_branch_locked_fitting}
\input{sections/appendices/04_hermite_interpolant}
\input{sections/appendices/05_fitting_coefficients}
\end{appendix}
\vfill
\clearpage

\bibliographystyle{unsrt}
\bibliography{bib/references}

\end{document}

%% file: settings/operator_definitions.tex
\newcommand{\nnn}{\nonumber\\}

\newcommand{\rnd}[1]{\!\left(#1\right)\xspace}
\newcommand{\sqr}[1]{\!\left[#1\right]\xspace}
\newcommand{\abs}[1]{\!\left|#1\right|\xspace}

\newcommand{\quotes}[1]{\lq#1\rq\xspace}

\newcommand{\eq}[1]{Eq.~\eqref{#1}\xspace}
\newcommand{\eqs}[2]{Eqs.~\eqref{#1}-\eqref{#2}\xspace}
\newcommand{\eqa}[2]{Eqs.~\eqref{#1} and \eqref{#2}\xspace}

\newcommand{\fig}[1]{Fig.~\ref{#1}\xspace}

\newcommand{\figa}[2]{Figs.~\ref{#1} and \ref{#2}\xspace}

\newcommand{\tab}[1]{Tab.~\ref{#1}\xspace}

\newcommand{\taba}[2]{Tabs.~\ref{#1} and \ref{#2}\xspace}

\newcommand{\app}[1]{App.~\ref{#1}\xspace}

%% file: settings/setup_options.tex
\usepackage{datetime}
\usepackage[utf8]{inputenc}
\usepackage[margin=0.75in]{geometry}
\usepackage{tcolorbox}
\usepackage[T1]{fontenc}

\usepackage{amsmath}
\usepackage{amssymb}
\usepackage{amsfonts}
\usepackage{graphicx}

\usepackage{xcolor}
\usepackage[english]{babel}
\usepackage[raggedright]{sidecap}
\usepackage[caption=false]{subfig}
\usepackage{ragged2e}
\usepackage{bm}
\usepackage{units}

\usepackage[sort&compress]{natbib}
\usepackage{xspace}
\usepackage{url}
\usepackage{tikz}
\usepackage{eso-pic}

\usepackage{placeins}
\usepackage{booktabs}
\usepackage{array}
\usepackage{multirow}
\usepackage{mathtools}
\usepackage{enumitem}
\usetikzlibrary{arrows.meta,positioning,shapes.geometric,fit,calc}

\usepackage{tikz}

\makeatletter
\newcommand\HUGE{\@setfontsize\Huge{30}{40}}
\makeatother

%% file: settings/tex_symbols.tex
\newcommand{\cvec}{\mathbf{c}}

\newcommand{\Evec}{\mathbf{E}}
\newcommand{\Fvec}{\mathbf{F}}

\newcommand{\Ccal}{\mathcal{C}}
\newcommand{\Dcal}{\mathcal{D}}
\newcommand{\Ecal}{\mathcal{E}}
\newcommand{\Fcal}{\mathcal{F}}
\newcommand{\Gcal}{\mathcal{G}}
\newcommand{\Hcal}{\mathcal{H}}
\newcommand{\HcalR}{\mathcal{H}_{\rm R}}
\newcommand{\Ical}{\mathcal{I}}
\newcommand{\Jcal}{\mathcal{J}}
\newcommand{\Kcal}{\mathcal{K}}

\newcommand{\Ocal}{\mathcal{O}}
\newcommand{\Pcal}{\mathcal{P}}
\newcommand{\Qcal}{\mathcal{Q}}
\newcommand{\Rcal}{\mathcal{R}}



%% file: settings/user_defined.tex
\newcommand{\dif}{\mathrm{d}\xspace}

\newcommand{\FD}[1]{\mathcal{F}_{#1}}
\newcommand{\HR}{\mathcal{H}_{\rm R}}
\newcommand{\HI}{\mathcal{H}_{\rm I}}
\newcommand{\kap}{\varkappa}

\NewDocumentCommand{\rectpulse}{r()}{%
	\mathord{\tikz[
		baseline=0pt,
		x=0.16em,
		y=1.6ex
		]{%
			\draw[
			line width=0.6pt,
			line cap=round
			]
			(0,0) -- (1,0) -- (1,1) --
			(5,1) -- (5,0) -- (6,0);
	}}\!\left(#1\right)%
}

%% file: sections/abstract.tex
An accurate analytic approximation is developed for the real boundary component of a regularised, causal complex auxiliary function describing the random phase approximation dielectric response of the weakly coupled electron gas across classical to strongly degenerate conditions. Its collisionless imaginary boundary component is known exactly from the Fermi--Dirac occupation logarithm. Following an assessment of existing approximations and the establishment of a high-precision numerical benchmark, a two-regime representation is constructed: a $6/7$ rational function at weak and moderate degeneracy and a composite polynomial approximation at strong degeneracy. Exact asymptotic constraints determine the rational denominator algebraically, leaving six independent state functions. Branch-locked optimisation, Chebyshev parametrisation and Bernstein-basis constraints provide accurate, smooth and numerically pole-free behaviour. In the strongly degenerate regime, the logarithmic Kohn contribution is isolated explicitly, with its regular remainder represented by analytically integrable polynomial expansions. The approximation achieves a maximum relative error of approximately $0.11\%$ in the real dielectric integral across the investigated domain, accurately reproduces its scaled derivative and is substantially faster than numerical quadrature. Validation using the electron--electron dynamic structure factor and imaginary-time correlation function yields maximum plasmon peak-amplitude errors of approximately $1.04\%$ and maximum pointwise correlation-function errors below $0.166\%$. Comparisons with established approximations demonstrate substantial improvements, particularly for weakly damped collective excitations. The resulting model provides an accurate, computationally efficient foundation for electronic screening, dynamic dielectric calculations and X-ray Thomson scattering analysis.

%% file: sections/main_body/01_introduction.tex
\section{Introduction}
\label{sec:introduction}

Electronic screening determines how a plasma responds to a charged perturbation and, consequently, the effective interactions that enter scattering, transport and energy relaxation. The familiar Debye--H\"{u}ckel and Thomas--Fermi descriptions specify its long-wavelength limit and are frequently used as the basis of simple calculations due to their straightforward and inexpensive analytic forms. However, integration over all momentum transfers is an essential aspect of many more precise theoretical models, requiring the electronic response at finite wave numbers. Finite-wavelength screening has been observed in quasi-elastic X-ray scattering from compressed matter \cite{Chapman_NatCommun_2015}, and the resulting screened ion potentials affect ionic structure predictions \cite{Moldabekov_ContribPlasmaPhys_2021}.

Recent wave number-resolved studies have reinforced the importance of electronic response in warm dense matter. Ab initio calculations for hydrogen find that electron localisation around ions changes
the static response at intermediate wave numbers \cite{Dornheim_MatterRadiatExtremes_2024}. X-ray scattering studies of compressed beryllium and aluminium have further exposed the sensitivity of inferred conditions and collective excitations to
electronic correlations and material structure \cite{Dornheim_NatCommun_2025,Bespalov_PhysRevLett_2026}.
These effects can require treatments beyond the uniform, non-interacting electron gas. Its dielectric response nevertheless provides a well-defined starting point whose complete wave number
dependence must itself be evaluated accurately.

In linear response, the dielectric screening function of the electronic subsystem is
\begin{align}
	\label{eq:Pi_ref}
	\epsilon_{ee}(k,\omega)
	= &\,
	1-\Pi_{ee}(k,\omega)V_{ee}(k)
	\,,
\end{align}
where $V_{ee}(k)=e^{2}/\epsilon_{0}k^{2}$ is the bare electron--electron Coulomb interaction and $\Pi_{ee}(k,\omega)$ is the
irreducible retarded density response to the total effective potential
\cite{Kremp_book}. In the random phase approximation (RPA), this is the retarded response of the ideal Fermi gas, which can be written in terms of a complex auxiliary function $\Hcal(z;\eta)$. The dynamic dielectric function on the real-frequency axis then takes the compact form
\begin{align}
    \label{eq:dynamic_dielectric_function}
    \epsilon_{ee}(k,\omega)
    = &\,
    1+\frac{\alpha^{2}}{2\varkappa}
    \sqr{
    	z_{+}\Hcal(z_{+};\eta)
    	-
    	z_{-}\Hcal(z_{-};\eta)
    }
    \, ,
\end{align}
The complex argument $z_{\pm} = x_{\pm} + i0^{+}$ has a positive infinitesimal imaginary part appropriate for the collisionless RPA form, and $x_{\pm}=w/\varkappa\pm\varkappa$ are the dimensionless variables indicating red- and blue-shifted energy and momentum transfer. The convention $\exp(-i\omega t)$ is used for the decay of adiabatically induced perturbations, such that the retarded continuation is analytic in the upper half of the Argand plane. The auxiliary function is defined by its regularised Cauchy transform,
\begin{align}
    \label{eq:H_complex_def}
    \Hcal(z;\eta)
    = &\,
    \frac{A(\eta)}{z}
    \sqr{
       \int_{-\infty}^{\infty}\!\dif x\,
       \frac{\FD{0}(\eta-x^{2})}{z-x}
	    +
	    i\pi\FD{0}(\eta)
    }
    \,
\end{align}
with $\mathfrak{Im}\{z\} > 0$ and $\FD{0}(y)=\ln(1+e^{y})$. The additive term in the square brackets removes a simple pole at the origin that otherwise arises from the Cauchy transform. In particular, the retarded function has a regular limiting value $\Hcal(0;\eta)=1$. The boundary values are
\begin{align}
    \label{eq:H_real_boundary}
    \HR(\chi;\eta)
    \equiv &\,
    \lim_{b\to0^{+}}\mathfrak{Re}\{\Hcal(\chi+ib;\eta)\}
    \,,
    \\
    \HI(\chi;\eta)
    \equiv &\,
    \lim_{b\to0^{+}}\mathfrak{Im}\{\Hcal(\chi+ib;\eta)\}
    \,.
\end{align}
\eq{eq:H_real_boundary} is the real function approximated throughout this paper. It has the equivalent nonsingular form
\begin{align}
    \label{eq:H_def}
    \HR(\chi;\eta)
    = &\,
    \frac{A(\eta)}{\chi}
    \int_0^\infty\frac{\dif x}{x}
    \ln\!\left[
       \frac{1+e^{\eta-(x-\chi)^2}}
            {1+e^{\eta-(x+\chi)^2}}
    \right]
    \, .
\end{align}
The imaginary boundary component of the regularised function is already known exactly from the Sokhotski--Plemelj formula:
\begin{align}
    \label{eq:H_exact_imag}
    \HI(\chi;\eta)
    = &\,
    \frac{\pi A(\eta)}{\chi}
    \sqr{\FD{0}(\eta)-\FD{0}(\eta-\chi^{2})}
    \, .
\end{align}
For clarity, the unregularised Cauchy function would instead have imaginary boundary value $-\pi A(\eta)\FD{0}(\eta - \chi^{2})/\chi$ and would be singular at $\chi=0$; the two representations produce exactly the same dielectric function because the regularisation constant cancels between the two terms of \eq{eq:dynamic_dielectric_function}. Thus, the new numerical model described in this paper approximates only $\HR$; the imaginary component in collisionless RPA is evaluated in closed form. In particular,
\begin{align}
    \label{eq:epsilon_exact_imag}
    \mathfrak{Im}\{\epsilon_{ee}(k,\omega)\}
    = &\,
    \frac{\pi A(\eta)\alpha^2}{2\varkappa}
    \ln\sqr{\frac{1 + e^{\eta - \rnd{w/\varkappa - \varkappa}^{2}}}{1 + e^{\eta - \rnd{w/\varkappa + \varkappa}^{2}}}}
    \,.
\end{align}

In the foregoing, $w=\hbar\omega/4k_{\rm B}T$,
$\varkappa=\hbar k/\sqrt{8m_{e}k_{\rm B}T}$, $\eta=\mu_{e}/k_{\rm B}T$,
$A(\eta)=\sqr{2\sqrt{\pi}\,\Fcal_{-1/2}(\eta)}^{-1}$ and
$\alpha=\kappa_{e}/k$ is the Salpeter collectivity parameter. 
The latter is familiar from characterising collective and non-collective regimes in optical/X-ray Thomson scattering
\cite{Evans_RepProgPhys_1969,Glenzer_RevModPhys_2009}. The inverse screening length used in this definition is suitable for arbitrary degeneracy
\begin{align}
	\label{eq:screening_wavenumber}
	\kappa_{e}^{2}
	= &\,
	\frac{e^{2}n_{e}}{\epsilon_{0}k_{\rm B}T}
	\frac{\Fcal_{-1/2}(\eta)}{\Fcal_{1/2}(\eta)}
	\,,
\end{align}
The Fermi--Dirac integrals (FDIs) are defined by
\begin{align}
	\label{eq:fdi_def}
	\Fcal_j(\eta)
	= &\,
	\frac{1}{\Gamma(j+1)}
	\int_0^\infty\dif t\,
	\frac{t^j}{e^{t-\eta}+1}
	\,.
\end{align}
These are defined for all orders $j > -1$ and extended to $j \leq -1$ by analytic continuation \cite{Kremp_book}.

Another important related function is the so-called loss function, defined as $L(k,\omega) = \mathfrak{Im}\{ -\epsilon_{ee}^{-1}(k,\omega) \}$. This has a single resonance (the plasmon pole) for $\alpha \gtrsim 1$, which occurs close to the weakly damped root of $\mathfrak{Re}\{\epsilon_{ee}(k,\omega)\}=0$. The detailed structure of the loss function peak is crucial to reconstruct in any fit of the dielectric function, especially for applications that directly integrate it, e.g., the electronic stopping power \cite{Dornheim_PhysRevB_2021}.

Because $\HR$ is even and the imaginary boundary component of $\Hcal$ is odd in $\chi$, the static limit of \eq{eq:dynamic_dielectric_function} is
\begin{align}
	\label{eq:static_dielectric_function}
	\epsilon_{ee}(k,0)
	= &\,
	1
	+
	\alpha^{2}
	\HR\rnd{\varkappa;\eta}
	\,.
\end{align}
The normalisation $\HR(0;\eta)=1$ gives
$\epsilon_{ee}(k,0) \approx 1+\kappa_{e}^{2}/k^{2}$ for $k\to0$.
This recovers Debye--H\"{u}ckel screening in the classical limit ($\eta \to -\infty$) and Thomas--Fermi screening in the degenerate limit ($\eta \to +\infty$). At fixed thermodynamic state, $\HR$ instead decays as $1/k^{2}$ for large $k$, giving $\epsilon_{ee}(k,0)-1\propto 1/k^{4}$. Neither specifies
the response where the momentum transfer is comparable to the inverse screening length, i.e., $\alpha \approx 1$. At
low temperature, the additional non-analytic structure near $k=2k_{\rm F}$ gives rise to Friedel oscillations \cite{Hagiya_PhysRevB_2020}.

For static processes, this finite-wavelength dependence enters through the screened electron--ion interaction, $V_{ei}^{\rm scr}(k) = V_{ei}(k)/\epsilon_{ee}(k,0)$ which impacts the strength of Rayleigh scattering \cite{Chapman_NatCommun_2015}, screened pseudopotentials for modelling structural properties \cite{Dharma-wardana_PhysRevE_2012}, provides the basic interaction governing local electron transport in Born approximation \cite{Ichimaru_PhysRevA_1985b, Apfelbaum_JPhysA_2006}, and modifies energy transfer between coupled collective modes \cite{Dharma-wardana_PhysRevLett_1991, Vorberger_PhysRevE_2010, Chapman_PhysRevE_2013}. Dynamic processes in which the screening function plays an important role include charged-particle stopping power \cite{Gericke_PhysRevE_2002b, BarrigaCarrasco_PhysRevE_2010}, local electron transport in the Lenard-Balescu picture \cite{Scullard_PhysPlasmas_2016, Veysman_PhysRevE_2016, Shaffer_PhysRevE_2020, Shaffer_PhysPlasmas_2024}, dynamical electron-ion collisional processes and laser-plasma coupling \cite{Millat_JPhysA_2003, Reinholz_PhysRevE_2000, Grinenko_PhysRevLett_2009, Reinholz_PhysRevE_2015} and predicting plasmon spectra in scattering experiments \cite{Glenzer_RevModPhys_2009}. For all such applications, numerical evaluation of $\HR$ is an efficiency bottleneck when repeated over many states, and especially from within nested momentum and/or frequency integrals.

The same ideal gas response also has uses beyond dielectric screening. Closely related integrals occur in the finite-temperature Hartree--Fock exchange self energy \cite{Kremp_book}; its static
spin response determines the finite-temperature RKKY interaction \cite{Kim_PhysRevB_1995}; and its wave number dependence supplies a linear-response constraint for nonlocal orbital-free free-energy functionals \cite{Sjostrom_PhysRevB_2013}. 

This work compiles the analytic limits of $\HR$, assesses existing approximations against numerical integration, and develops an accurate rational representation for repeated evaluation. The accuracy, denominator stability and execution time are compared to identify a practical method for calculations that require the full RPA response. In the present collisionless study, the fitted real boundary is combined with the exact imaginary boundary expression \eqref{eq:H_exact_imag}; construction of an approximately causal continuation of the fitted boundary into the upper complex plane is reserved for subsequent work.

%% file: sections/main_body/02_analytic_properties.tex
\section{Analytic properties of $\HR$}

\subsection{Alternative integral forms}

For the real boundary value $\HcalR$ defined in \eq{eq:H_real_boundary}, the principal-value integral \eqref{eq:H_def} has several equivalent representations that are of use to this work. Each type is advantageous for the derivation of specific analytic results in limiting cases. First, it is possible to readily manipulate \eq{eq:H_def} into the Hilbert transform of the zeroth-order FDI:
\begin{align}
	\label{eq:H_def_2}
	\HcalR(\chi;\eta)
	= &\,
	\frac{A(\eta)}{\chi}
	\Pcal
	\int_{-\infty}^{\infty}
	\dif x\,
	\frac{\FD{0}(\eta - x^{2})}{\chi - x}
	\,,
\end{align}
The symbol $\Pcal$ stands for the Cauchy principal value. The resulting real-axis function is smooth at finite $\eta$ and real $\chi$, but it is not itself the holomorphic continuation to complex arguments. In this form, $\HcalR$ is clearly a degenerate generalisation of the Hilbert transform of the Gaussian function $\exp(\eta - x^{2})$, which was extensively discussed by Fried and Conte \cite{Fried_book}. Indeed, as discussed here it is clearly nothing more than a rescaled version of the real part of the generalisation introduced more recently by Melrose and Mushtaq \cite{Melrose_PhysPlasmas_2010}.

Two other useful equivalent forms of $\HcalR$ can be obtained by applying successive integration by parts to \eq{eq:H_def_2}. The result can be shown to be \cite{Arista_PhysRevA_1984}
\begin{align}
	\label{eq:H_def_3}
	\HcalR(\chi;\eta)
	= &\,
	\frac{A(\eta)}{\chi}
	\int_{0}^{\infty}
	\dif x\,
	\frac{2x}{1 + e^{x^{2} - \eta}}
	\ln\abs{\frac{x + \chi}{x - \chi}}
	\,.
\end{align}
A second integration by parts then leads to \cite{Ancarani_EurPhysJPlus_2016}
\begin{align}
	\label{eq:H_def_4}
	\HcalR(\chi;\eta)
	= &\,
	\frac{A(\eta)}{\chi}
	\int_{0}^{\infty}
	\dif x \, x^{3}\, \text{sech}^{2}\rnd{\frac{x^{2} - \eta}{2}}
	g\rnd{\frac{\chi}{x}}
	\,,
	\\
	\label{eq:G_def}
	g\rnd{x}
	= &\,
	x
	+
	\frac{1}{2}
	\rnd{1 - x^{2}}
	\ln\abs{\frac{1 + x}{1 - x}}
	\,.
\end{align}
Here, $g(x)$ is the Lindhard function \cite{Lindhard_MatFysMedd_1954}, which is well known from the literature for the description of screening in solid state systems.

Considering the point $x = 1$, the Lindhard function \eq{eq:G_def} remains finite with value $g(1) = 1$ but its derivative is singular $g'(1) = -\infty$. This is the so-called \quotes{Kohn anomaly}, where the derivative shows a logarithmic divergence \cite{Kohn_PhysRevLett_1959}. Capturing the emergence of this behaviour from an otherwise well-behaved and smooth function is the most difficult challenge constructing in a global approximation for $\HcalR$.

As will be discussed, the derivative of $\HcalR$ is also of interest; throughout this paper $\Ical$ denotes the real-axis scaled derivative $-\chi\,\partial\HcalR/\partial\chi$. From \eq{eq:H_def_4} one finds
\begin{align}
	\label{eq:H_deriv}
	\Ical(\chi;\eta)
	= &\,
	\frac{A(\eta)}{\chi}
	\int_{0}^{\infty}
	\dif x\,
	x^{2} \mathrm{sech}^{2}\rnd{\frac{x^{2} - \eta}{2}}
	\frac{x^{2}}{\chi^{2} - x^{2}}
	\nnn & 
	\times
	\sqr{\frac{2\chi}{x} - \rnd{1 + \frac{\chi^{2}}{x^{2}}}g\rnd{\frac{\chi}{x}}}
	\,.
\end{align}
The behaviour of this function at the Kohn point remains integrable despite diverging logarithmically in the quantum limit.

\subsection{Expansions in $\varkappa$ for arbitrary $\eta$}

As mentioned earlier, the principal role of $\HcalR$ is to describe the electronic screening in plasma. In the static case, it extends the commonly used long-wavelength ($\varkappa \to 0$) Debye-H\"{u}ckel form to arbitrary wave number shifts (finite wavelengths). This should therefore be accessible directly from Taylor expansion of $\HcalR$ about $\varkappa = 0$. Similarly, analytic results are known in the short-wavelength ($\varkappa \to \infty$) limit. Both are best understood as complementary expansions of the Hilbert transform kernel $1/(x - \chi)$ in \eq{eq:H_def_2}.

\subsubsection{Long-wavelength limit: $\HcalR(\kap \to 0; \eta)$}

Here the appropriate expansion is obtained from
$$
\frac{1}{x - \kap} 
= \frac{1}{x}\frac{1}{1 - \kap/x}
$$
and expanding the second fraction for small $\kap/x$, giving 
\begin{align}
	\label{eq:kap_to_0_expansion}
	\lim_{\kap\to0}
	\frac{1}{x - \kap}
	=
	\frac{1}{x}\sqr{1 + \frac{\kap}{x} + \frac{\kap^{2}}{x^{2}} + \dots}
	\,.
\end{align}
Plugging \eq{eq:kap_to_0_expansion} into \eq{eq:H_def_2} gives 
\begin{align}
	\HcalR(\kap;\eta)
	= &\,
	\frac{A(\eta)}{\varkappa}\,
	\Pcal
	\int_{-\infty}^{\infty}
	\frac{\dif x}{x}
	\sqr{1 + \frac{\kap}{x} + \frac{\kap^{2}}{x^{2}} + \dots}
	\nnn & 
	\times
	\ln\rnd{1 + e^{\eta - x^{2}}}
	\,,
\end{align}
which gives the general series result
\begin{align}
	\label{eq:H_kap_to_0_expansion}
	\HcalR(\kap;\eta)
	= &\,
	\sum_{m = 0}^{\infty}
	\frac{(-2)^{m}}{(2m + 1)!!}
	\frac{\Fcal_{-m-1/2}(\eta)}{\Fcal_{-1/2}(\eta)}
	\kap^{2m}
	\,.
\end{align}
The first few terms of this series give
\begin{align}
	\label{eq:smallk}
	\HcalR(\kap;\eta)
	\sim &\,
	1
	-
	\frac{2}{3}\frac{\FD{-3/2}(\eta)}{\FD{-1/2}(\eta)}\kap^2
	+
	\frac{4}{15}\frac{\FD{-5/2}(\eta)}{\FD{-1/2}(\eta)}\kap^4
	\nnn &
	-
	\frac{8}{105}\frac{\FD{-7/2}(\eta)}{\FD{-1/2}(\eta)}\kap^6
	+
	O(\kap^8)
	\,.
\end{align}
The exact leading order result is therefore
\begin{equation}
	\label{eq:H_zero_kap}
	\lim_{\varkappa\to0}\HcalR(\varkappa;\eta) = 1
\end{equation}
for any finite level of degeneracy, i.e., any value of $\eta$. Subsequently, the long-wavelength limit of the static dielectric function is not modified from the Debye-H\"{u}ckel/Thomas-Fermi forms, as required.

\subsubsection{Short-wavelength limit: $\HcalR(\varkappa \to \infty,\eta)$}

In the opposite short-wavelength limit the appropriate expansion is obtained by writing
$$
\frac{1}{x - \kap} = -\frac{1}{\kap}\frac{1}{1 - x/\kap}
$$
and expanding the second fraction for small $x/\kap$, giving the complementary series
\begin{align}
	\label{eq:kap_to_infty_expansion}
	\lim_{\kap\to\infty}
	\frac{1}{x - \kap}
	=
	\frac{1}{\kap}\sqr{1 + \frac{x}{\kap} + \frac{x^{2}}{\kap^{2}} + \dots}
	\,.
\end{align}
We therefore find for the $\HcalR$-function
\begin{align}
	\HcalR(\kap;\eta)
	= &\,
	\frac{A(\eta)}{\kap^{2}}\,
	\Pcal
	\int_{-\infty}^{\infty}
	\dif x
	\sqr{1 + \frac{x}{\kap} + \frac{x^{2}}{\kap^{2}} + \dots}
	\nnn & 
	\times
	\ln\rnd{1 + e^{\eta - x^{2}}}
	\,,
\end{align}
with the general result equivalent to \eq{eq:H_kap_to_0_expansion}
\begin{align}
	\label{eq:H_kap_to_infty_expansion}
	\HcalR(\kap;\eta)
	= &\,
	\sum_{m = 0}^{\infty}
	\frac{(2m - 1)!!}{2^{m+1}}
	\frac{\Fcal_{m+1/2}(\eta)}{\Fcal_{-1/2}(\eta)}
	\kap^{-2m - 2}
	\,.
\end{align}
For any $\eta$ and $\kap\rightarrow\infty$, we find the equivalent to \eq{eq:smallk}
\begin{align}
	\label{eq:largek}
	\HcalR(\kap;\eta)
	\sim &\,
	\frac{1}{2}\frac{\FD{1/2}(\eta)}{\FD{-1/2}(\eta)}\frac{1}{\kap^2}
	+
	\frac{1}{4}\frac{\FD{3/2}(\eta)}{\FD{-1/2}(\eta)}\frac{1}{\kap^4}
	\nnn &
	+
	\frac{3}{8}\frac{\FD{5/2}(\eta)}{\FD{-1/2}(\eta)}\frac{1}{\kap^6}
	+O(\kap^{-8})
	\,.
\end{align}

For extremely large values of $\kap$ one finds
\begin{align}
	\label{eq:H_infty_kap}
	\lim_{\varkappa\to\infty}
	\kap^2\HR(\kap;\eta)
	= &\,
	\frac{1}{2}\frac{\Fcal_{1/2}(\eta)}{\Fcal_{-1/2}(\eta)}
	\,.
\end{align}
It should be noted that technically the binomial expansion of the singular term in the general definition is not strictly allowed, but converges properly in this case due to the exponential modulation of the logarithmic term.

\subsection{Expansions in $\eta$ for arbitrary $\kap$}

Expanding $\HcalR$ instead about the degeneracy variable $\eta$ into the classical ($\eta \to -\infty$) and quantum ($\eta \to +\infty$) regimes whilst considering arbitrary values of the dimensionless wave number $\kap$ is also amenable to analytic series with well-known leading-order behaviours.

\subsubsection{Classical limit: $\HcalR(\varkappa;\eta \to -\infty)$}

In the classical limit the ratio of any two normalised FDIs is precisely unity, since $\Fcal_{j}(\eta \to -\infty) = e^{\eta}$ for all $j$. The result for arbitrary $\kap$ is straightforwardly obtained from the Hilbert transform form of $\HcalR$ \eqref{eq:H_def_2} noting that for $\eta \to -\infty$ the FDI of order $j = 0$ simply becomes a scaled Gaussian, i.e., 
\begin{align}
	\label{eq:F0_classical}
	\FD{0}(\eta - x^{2})
	= &\,
	\sum_{s = 1}^{\infty}
	(-1)^{s + 1}\frac{e^{s\eta}}{s}e^{-sx^{2}}
	\approx
	e^{\eta}e^{-x^{2}}
	\,.
\end{align}
Denoting $\xi = e^{\eta}$, then this is precisely the fugacity expansion considered by Melrose and Mushtaq. Taking the leading term only (the classical limit) and substituting into \eq{eq:H_def_2} gives \cite{Tsytovich_AstropartPhys_1996}
\begin{align}
	\label{eq:H_classical}
	\HcalR(\varkappa)
	= &\,
	-\frac{1}{2\sqrt{\pi}\kap}\,
	\Pcal
	\int_{-\infty}^{\infty}
	\frac{\dif x}{x - \varkappa}\,
	e^{-x^{2}}
	\nnn
	= &\,
	\frac{\Dcal(\kap)}{\kap}
	\,,
\end{align}
where $\Dcal(x)$ is known as the Dawson integral of the first kind \cite{Abramowitz_book}, which can be accurately represented using the rational minimax approach of Roy \cite{Roy_CompPhysCommun_2009}. Expansion of this in the limits $\kap\to0$ and $\kap\to\infty$ give
\begin{align}
	\label{eq:H_classical_smallk}
	\HcalR(\varkappa)
	\sim &\,
	1
	-
	\frac{2}{3}\kap^2
	+
	\frac{4}{15}\kap^4
	-
	\frac{8}{105}\kap^6
	+
	O(\kap^8)
	\,,
	\\
	\label{eq:H_classical_largek}
	\HcalR(\varkappa)
	\sim &\,
	\frac{1}{2\kap^2}
	+
	\frac{1}{4\kap^4}
	+
	\frac{3}{8\kap^6}
	+O(\kap^{-8})
	\,.
\end{align}

\subsubsection{Quantum limit: $\HcalR(\varkappa;\eta \to \infty)$}

In the deeply degenerate regime the ratio of FDIs instead tends to a power law in the degeneracy variable
$$
\frac{\FD{j}(\eta\to\infty)}{\FD{k}(\eta\to\infty)}
=
\frac{\Gamma(k + 2)}{\Gamma(j + 2)} \eta^{j - k}
\,.
$$
The complementary expansion to \eq{eq:F0_classical} is obtained by writing $\FD{0}(\eta - x^{2}) = \ln(e^{\eta - x^{2}}(1 + e^{x^{2} - \eta})) = \eta - x^{2} + \FD{0}(x^{2} - \eta)$ and applying the logarithmic expansion to the second term. Importantly, this only applies when $\eta > x^{2}$, and only the leading order makes any noteworthy contribution, since all other terms are exponentially small. The leading order approximation to the relevant function then becomes
\begin{align}
	\label{eq:F0_degenerate}
	\FD{0}(\eta - x^{2}; \eta \to \infty)
	\approx
	\rnd{\eta - x^{2}}
	\Theta\rnd{\eta - x^{2}}
	\,,
\end{align}
where $\Theta(x)$ is the Heaviside step function. 
Here, a more natural position from which to derive the required result is to start 
\eq{eq:H_def_4}. Since the sech term has the limiting behaviour $\lim_{\eta \to \infty}\frac{1}{4}\mathrm{sech}^{2}(\frac{1}{2}(x^{2} - \eta)) = \delta(x^{2} - \eta)$, then by defining $y = x^{2}/\eta$ \eq{eq:H_def_4} straightforwardly yields
\begin{align}
	\label{eq:H_quantum}
	\HcalR(\varkappa_{0})
	= &\,
	\frac{1}{2\varkappa_{0}}
	\int_{0}^{\infty}
	\dif y \, y\, \delta\rnd{y - 1}
	g\rnd{\frac{\varkappa_{0}}{\sqrt{y}}}
	\nnn 
	= &\,
	\frac{g\rnd{\varkappa_{0}}}{2\varkappa_{0}}
	\,,
\end{align}
with $\varkappa_{0} = \varkappa/\sqrt{\eta}$. The small and large $\varkappa$ expansions of this result give
\begin{align}
	\label{eq:H_quantum_smallk}
	\HcalR(\varkappa_{0})
	\sim &\,
	1
	-
	\frac{\varkappa_{0}^{2}}{3}
	-
	\frac{\varkappa_{0}^{4}}{15}
	-
	\frac{\varkappa_{0}^{6}}{35}
	+
	O(\varkappa_{0}^8)
	\,.
	\\
	\label{eq:H_quantum_largek}
	\HcalR(\varkappa_{0})
	\sim &\,
	\frac{1}{3\varkappa_{0}^{2}}
	+
	\frac{1}{15\varkappa_{0}^{4}}
	+
	\frac{1}{35\varkappa_{0}^{6}}
	+
	O(\varkappa_{0}^{-8})
	\,.
\end{align}

%% file: sections/main_body/03_numerical_benchmark.tex
\section{Numerical benchmark}
\label{sec:numref}

Given that the aim of this work is to provide an efficient and accurate approximation to $\HcalR$, it is first necessary to give a credible and independent reference case, which must be achieved using numerical integration. It is of interest to also find a maximally efficient approach to this benchmark result, which is established by applying a common algorithm and enabling variable transform to deal with the improper limit(s) of integration required. The three cases considered are: \eq{eq:H_def}, \eq{eq:H_def_3} and \eq{eq:H_def_4}, from which we construct an optimal approach spanning a wide range of the independent variable $\chi$ and the parameter $\eta$. The Hilbert form \eqref{eq:H_def_2} is not considered due to the additional overhead associated with handling the simple pole; the other forms are more fairly matched and can all be handled with the same numerical scheme. 

For each of the integral forms of interest, the improper limits of integration must be handled. For any uniform integration variable distribution this is clearly infeasible, although clearly only some finite upper limit that captures the majority of the integrand structure needs to be considered. Since the structure changes over the $\chi$--$\eta$ plane, however, it is more convenient to compactify the integrand onto a universal domain. This can be achieved using a number of different transformations of integration variable, all of which compress the portion of the original space into a small, finite portion of the transformed space. 

A suitable choice for this application is the tangent transform: $x(u) = \tan(\frac{\pi}{2}u)$, and $\dif x = \frac{\pi}{2}\sec^{2}(\frac{\pi}{2}u) \dif u = \frac{\pi}{2}(1 + x(u)^{2}) \dif u$. The integration domain then changes from $x \in [0, \infty)$ to $u \in [0, 1)$. With this transformation the three integrals of interest are:
\begin{align}
	\label{eq:H_transformed}
	\Hcal_{n}(\chi;\eta)
	= &\,
	\frac{\pi A(\eta)}{2\chi}
	\int_{0}^{1}
	\dif u\,
	\rnd{1 + x^{2}(u)}
	\nnn
	& \times
	\Ccal_{n}(x(u), \chi; \eta)
	\,,
	\\
	\label{eq:K_1_def}
	\Ccal_{\rm I}(x, \chi; \eta)
	= &\,
	\frac{1}{x}
	\ln\sqr{\frac{1 + e^{\eta - (x - \chi)^{2}}}{1 + e^{\eta - (x + \chi)^{2}}}}
	\,,
	\\
	\label{eq:K_2_def}
	\Ccal_{\rm II}(x, \chi; \eta)
	= &\,
	\frac{2x}{1 + e^{x^{2} - \eta}}
	\ln\abs{\frac{x + \chi}{x - \chi}}
	\,,
	\\
	\label{eq:K_3_def}
	\Ccal_{\rm III}(x, \chi; \eta)
	= &\,
	x^{3}\,
	\mathrm{sech}^{2}
	\rnd{\frac{x^{2} - \eta}{2}}
	g\rnd{\frac{\chi}{x}}
	\,.
\end{align}
An open quadrature method is appropriate in all cases since the integration limits themselves are not used.

\begin{figure}[!t]
	\centering
	\includegraphics[width=0.49\textwidth]
	{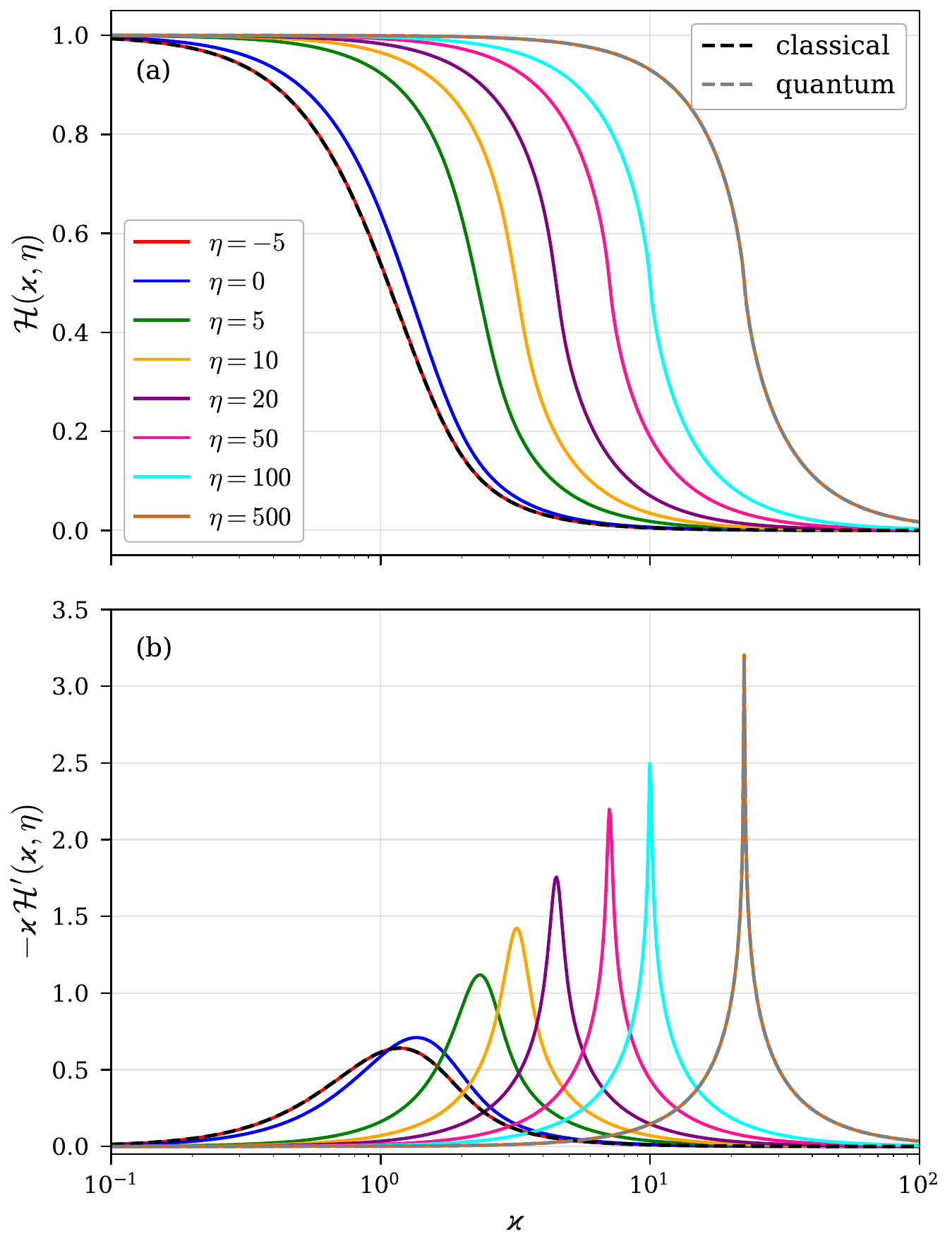}
	\caption{
		Variation of $\HcalR$ (a) and $\Ical$ (b) as a function of $\chi$ for a range of $\eta$ values using the Type III numerical integration baseline. The classical result is shown in both panels with the black dashed curves and the purely quantum result are shown by the dashed grey curves.
	}
	\label{fig:numdirect_numtan_H}
\end{figure}

For the numerical baseline the transformed integrals \eq{eq:H_transformed} are evaluated using the globally adaptive QUADPACK algorithm implemented in \texttt{scipy.integrate.quad} \cite{Piessens_book}. On each subinterval the method employs an embedded 21-point Gauss–Kronrod quadrature rule, with adaptive subdivision driven by the estimated local errors and convergence acceleration by extrapolation. A global relative error of $10^{-10}$ is required for the baseline, so that accuracy of the various approximate forms can be robustly compared.

The results of numerical integration are shown in \fig{fig:numdirect_numtan_H}. In Panel (a), the variation of the $\HcalR$ is shown as a function of $\chi$ for a wide range of $\eta$ values spanning the space between the classical (black dashed curve) and quantum (grey dashed curve) limits. Panel (b) shows the negative scaled derivative $\Ical$. The results shown use the Type III form of the integration, which is the most numerically stable form as it does not have to fight precision issues at large $\chi$ (Type I) or resolve the integrable logarithmic singularity (Type II). The Type III form of the integrand is therefore chosen as the baseline for the remainder of this work.

As shown in \fig{fig:numdirect_numtan_H} the limiting forms for the classical and quantum regimes \eqa{eq:H_classical}{eq:H_quantum}, respectively, are indistinguishable from the numerical integration results for sufficiently large negative/positive values of $\eta$. The values at which the limiting forms \eqa{eq:H_classical}{eq:H_quantum} can be established by root-solving for $\eta$ such that the maximum absolute relative error ($L_{\infty}$) over $\varkappa$ is below some small threshold value; these are denoted $\eta_{\rm class}(L_{\infty})$ and $\eta_{\rm quant}(L_{\infty})$. In the classical regime a cross-over point of $\eta_{\rm class}(10^{-6}) \approx -13$ is readily found. In the quantum limit such a criterion is too stringent due to the shape of the Kohn point in the quantum regime. It is therefore relaxed to $L_{\infty} \leq 10^{-3}$ for the quantum regime, from which a value $\eta_{\rm quant}(10^{-3}) \approx 478$ is obtained. The classical cross-over with this same relaxed threshold is $\eta_{\rm class}(10^{-3}) \approx -6$.

%% file: sections/main_body/04_existing_approximations.tex
\section{Existing global approximations}

With a robust and efficient numerical benchmark established it is now possible to quantify the accuracy and performance of the various approximate forms available in the literature, as well as the new developments made in this work.

\subsection{Dandrea-Ashcroft-Carlsson 4/5 rational approximation}

The first approximate form considered is inspired by the work of Dandrea, Ashcroft and Carlsson (DAC), \cite{Dandrea_PhysRevB_1986}, who constructed a closed-form approximation to the finite-temperature, frequency-dependent non-interacting dynamic electron polarisation function. Their primary fitting variables are the dimensionless frequency and wave number variables
\begin{align}
	\label{eq:Qkappa}
	z 
	= &\,
	\frac{\hbar\omega}{4k_{\rm B}T_{\rm F}}
	\,,
	~~~
	q
	=
	\frac{k}{2k_{\rm F}}
	=
	\sqrt\Theta\,\kap
	\,,
\end{align}
in which $T_{\rm F} = \hbar^{2}k_{\rm F}^{2}/2m_{e}k_{\rm B}$ and $k_{\rm F} = (3\pi^{2}n_{e})^{1/3}$ are the Fermi temperature and wave number, respectively, and $\Theta = T/T_{\rm F}$ is the degeneracy parameter. For the dynamic problem the wave-number and frequency dependence are folded into the single argument $x=z/q \pm q$; for the static problem of interest here $x \equiv q$. DAC then fit a single rational function $P(x;\Theta)$ \cite{Dandrea_PhysRevB_1986}. Their construction was expressly motivated by computational efficiency: the paper reports that use of the rational form in place of the defining integral reduced the cost of a representative free-energy calculation by about two orders of magnitude while changing the result by at most about two percent.

The fitted rational function given by DAC is
\begin{align}
	\label{eq:dacP}
	P(x;\Theta)
	= &\,
	\frac{1 + \displaystyle\sum_{j=1}^{4}a_{2j}(\Theta)x^{2j}}
	{1 + \displaystyle\sum_{k=1}^{5}b_{2k}(\Theta)x^{2k}}
	\,,
\end{align}
We can then explicitly identify $\HcalR$ with the above, i.e.,	
\begin{align}
	\label{eq:dac}
	\HR^{\rm DAC}(\chi;\eta) 
	\equiv &\,
	P(x;\Theta)
	\,.
\end{align}
The four numerator functions $a_{2j}(\Theta)$ are empirical fits in the parameter $\Theta$ (see \app{app:coefficients}). 

The denominator is not fitted independently. DAC constrain it using the small-$x$ and large-$x$ behaviours of the exact response, as previously given in \eq{eq:smallk} and \eq{eq:largek}. They work with the unnormalised form of the FDIs: $I_j(\eta) = \Gamma(j+1)\FD{j}(\eta)$, which are also approximated as rational functions of $\Theta$ to give $I_{j}(\Theta)$. The construction of the coefficients of $P(x,\Theta)$ is then
\begin{align}
	\label{eq:dacblo}
	b_{10}(\Theta)
	= &\,
	\frac32\sqrt\Theta\,
	I_{-1/2}(\Theta)a_8(\Theta)
	\,,
	\nnn
	b_8(\Theta)
	= &\,
	\sqrt\Theta
	\left[
		\frac32I_{-1/2}(\Theta)a_6(\Theta)
		-\frac12\Theta^2I_{3/2}(\Theta)b_{10}(\Theta)
	\right]
	\,,
	\nnn
	b_6(\Theta)
	= &\,
	\sqrt\Theta
	\left[
		\frac32I_{-1/2}(\Theta)a_4(\Theta)
		-
		\frac12\Theta^2I_{3/2}(\Theta)b_8(\Theta)
		\right. \nnn &\left. 
		-
		\frac{3}{10}\Theta^3I_{5/2}(\Theta)b_{10}(\Theta)
	\right]
	\,,
	\\
	\label{eq:dacbhi}
	b_2(\Theta)
	= &\,
	a_2(\Theta)
	+
	\frac{2J(\Theta)}{3\sqrt\Theta\,I_{-1/2}(\Theta)}
	\,,
	\nnn
	b_4(\Theta)
	= &\,
	b_2(\Theta)^2
	-
	a_2(\Theta)b_2(\Theta)
	+
	a_4(\Theta)
	\nnn &
	+
	\frac{2K(\Theta)}{15\sqrt\Theta\,I_{-1/2}(\Theta)}
	\,.
\end{align}
The two auxiliary quantities $J$ and $K$ are approximated with further Pad\'{e} formulae in $\Theta$, as well as the required FDIs $I_{j}(\Theta)$. The full set of equations used by the model is described in \app{app:DAC_model_equtions} and the complete coefficient set is tabulated in \tab{tab:original_DAC_coefficients}. 

The relative errors of DAC's approximation scheme using their coefficients $a_{j}(\Theta)$ and $b_{j}(\Theta)$ as per \eqs{eq:dacblo}{eq:dacbhi} compared to the robust Type III numerical integration results yield a maximum error of $3.54\%$, which occurs at about $\chi = 23.41$ for $\eta = 500$. The $L_{1}$ and $L_{2}$ errors are $0.27\%$ and $0.66\%$, respectively. Of course these numbers belie the impact of the singular behaviour in DAC's Pad\'e constructions; those points are not included in the array of $\chi$ and $\eta$ values considered.

\subsection{Hybrid/re-optimised DAC 4/5 rational approximation}

Although the original implementation of the DAC model is simple and reasonably accurate, there are some simple ways in which it can be improved. First, the various FDIs can be replaced with the Fukushima-minimax implementation \cite{Fukushima_ApplMathComput_2015, Fukushima_ApplMathComput_2015b} as distributed in Maddox's \texttt{fdint} package \cite{MaddoxFDINT}. This change enables floating point accuracy at speeds comparable to that of most standard functions. Second, it is possible to completely eliminate the rational functions $J(\Theta)$ and $K(\Theta)$, since these can be manipulated into negative-half-order FDIs
\begin{align}
	\label{eq:Jfdi}
	J(\Theta,\eta)
	= &\,
	\sqrt{\frac{\pi}{\Theta}}\,
	\FD{-3/2}(\eta)
	\,,
	\\
	\label{eq:Kfdi}
	K(\Theta,\eta)
	= &\,
	-\frac{2}{\Theta}
	\sqrt{\frac{\pi}{\Theta}}\,
	\FD{-5/2}(\eta)
	\,.
\end{align}
In particular, the above substitution removes a singularity in $J(\Theta)$ that occurs due to a zero in its denominator around $\Theta \approx 0.251606447$, i.e., for $\eta \approx 3.71431$, and improves the corresponding structure in $K(\Theta)$.

\begin{figure}[!t]
	\includegraphics[width=0.48\textwidth]
	{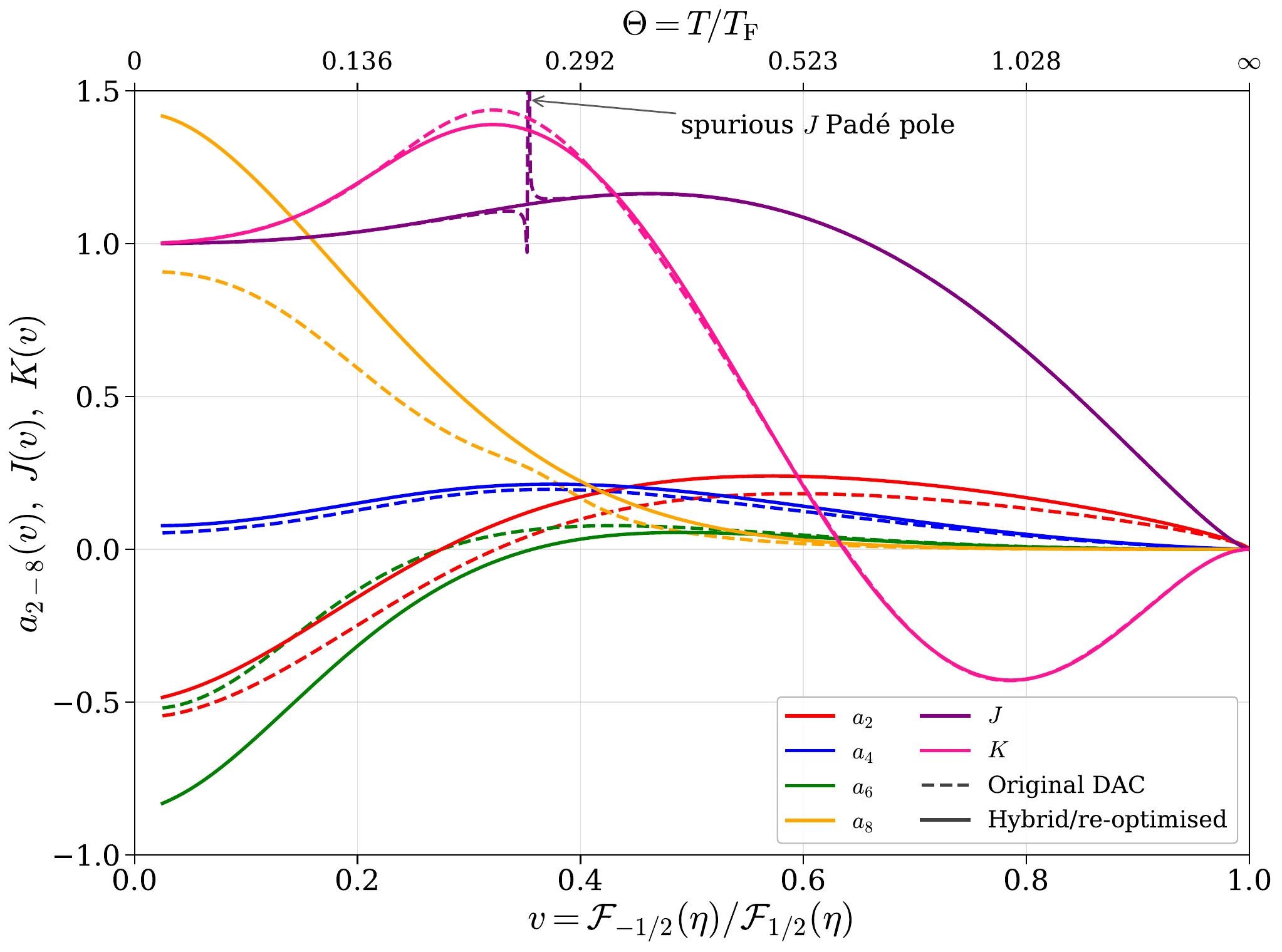}
	\caption{
		Coefficient functions entering the DAC construction plotted against $v = \FD{-1/2}(\eta) / \FD{1/2}(\eta)$. Dashed curves use the original DAC implementation and solid curves the first FLF overhaul using Fukushima-minimax FDIs.  The $a_{2j}$ curves coincide by construction. Dense sampling exposes the narrow spurious pole in the published fit for $J(\Theta)$ near $v=0.353$, while $K(\Theta)$ shows a broader discrepancy which is driven by limited accuracy fitting rather than singular behaviour.
	}
	\label{fig:daccoeff}
	\vspace{0.5cm}
\end{figure}

The original (dashed curves) and improved (solid curves) forms of the coefficients $b_{2-10}(\Theta,\eta)$,  $J(\Theta,\eta)$ and $K(\Theta,\eta)$ are shown in \fig{fig:daccoeff}. The singularity and poor accuracy in the now-unnecessary terms are clearly visible. Note that the comparison has been performed in a new compactified coordinate
\begin{align}
	\label{eq:v_def}
	v(\eta)
	\equiv &\,
	\Gcal_{-1/2}(\eta)
	\,,
	~~~
	\Gcal_{j}(\eta)
	\equiv
	\frac{\FD{j}(\eta)}{\FD{1/2}(\eta)}
	\,,
\end{align}
which compresses the whole range $\Theta \in (0, \infty)$ to the well-defined range $v \in (0, 1)$. 

Having eliminated all of the low-accuracy rational functions other than the numerator coefficients $a_{2j}(\Theta)$ from the improved DAC fit, some indication of two other avenues for improvement become clear. The first, and most obvious, is that it may be far easier to parametrise the entire system of equations in terms of the compactified coordinate $v(\eta)$ defined in \eq{eq:v_def}. Second, the existing rational fits in $\Theta$ are entirely responsible for the reduced accuracy in the transition regime between the classical and quantum limiting forms. Hints of this can be seen in \fig{fig:daccoeff}, where some unexpected oscillatory behaviour can be seen; see $a_{8}(\Theta)$ around $v \sim 0.3$. 

Noting the relationship between the chemical potential and the usual degeneracy parameter
\begin{align}
	\label{eq:Theta_def}
	\Theta(\eta)
	= &\,
	\rnd{\frac{4}{3\sqrt{\pi}\Fcal_{1/2}(\eta)}}^{2/3}
	\,,
\end{align}
then the coefficients $b_{j}$ can be written as
\begin{align}
	b_{10}(v, \eta)
	= &\,
	\frac{3\sqrt{\pi}}{2}\,
	\Theta(\eta)^{1/2}
	\Fcal_{1/2}(\eta)\,
	v\,a_8(v)
	\,,
	\nnn
	b_8(v, \eta)
	= &\,
	\frac{3\sqrt{\pi}}{2}
	\Theta(\eta)^{1/2}
	\Fcal_{1/2}(\eta)
	\nnn & 
	\times
	\left[
		v\,a_6(v)
		- 
		\frac{1}{4}\Theta(\eta)^2\Gcal_{3/2}(\eta)
		b_{10}(v,\eta)
	\right]
	\,,
	\nnn
	b_6(v,\eta)
	= &\,
	\frac{3\sqrt{\pi}}{2}
	\Theta(\eta)^{1/2}
	\Fcal_{1/2}(\eta)
	\nnn &
	\times
	\left[
		v\,a_4(v)
		-
		\frac{1}{4}\Theta(\eta)^2\Gcal_{3/2}(\eta)
		b_8(v,\eta)
		\right. \nnn & \left. 
		-
		\frac{3}{8}\Theta(\eta)^3\Gcal_{5/2}(\eta)
		b_{10}(v,\eta)
	\right]
	\,,
	\\
	b_2(v, \eta)
	= &\,
	a_2(v)
	+
	\frac{2}{3}
	\frac{\Gcal_{-3/2}(\eta)}{v\,\Theta(\eta)}
	\,,
	\nnn
	b_4(v, \eta)
	= &\,
	b_2^2(v, \eta)
	-
	a_2(v)b_2(v, \eta)
	+
	a_4(v)
	\nnn &
	-
	\frac{4}{15}
	\frac{\Gcal_{-5/2}(\eta)}{v\,\Theta(\eta)^2}
	\,,
\end{align} 
for a new set of re-optimised coefficients $a_{j}(v)$.

Another strong reason for changing the fitting variable from $\Theta$ to $v$ is that the functions to fit exist on the interval $(0, 1)$, which enables the use of Bernstein polynomials to ensure positivity of the denominators of the rational forms of the new $a_{j}(v)$. Assuming the same 4/4 rational function form the coefficients are chosen to be
\begin{align}
	\label{ewq:new_a_2j_form}
	a_{j}(v)
	= &\,
	\frac{\displaystyle\sum_{m = 0}^{4}p_{jm}\rnd{1 - v}^{(2m + j)/3}}{1 + \displaystyle\sum_{m = 1}^{4}q_{jm}\rnd{1 - v}^{2m/3}}
	\,.
\end{align}
The variable $\rnd{1 - v}^{2/3}$ is chosen since it has the asymptotic behaviour $\rnd{1 - v}^{2/3} \propto 1/\Theta$ as $v \to 1$. 	

A re-optimisation of the numerator coefficients has been performed using $100$ evenly spaced samples in $v \in (0, 1)$, resulting in new coefficients (\tab{tab:hybrid_DAC_coefficients}). The independent state-wise points are generated by fitting the hybrid 4/5 rational fit against the Type-III numerical benchmark results. The coefficients are broadly similar to those of DAC, although $a_{6}$ and $a_{8}$ differ substantially in the $v \to 0$ limit (corresponding to the quantum regime $\eta \to +\infty$). In terms of error, the re-optimised results have slightly better $L_{1}$ and $L_{2}$ values: $0.24\%$ and $0.58\%$, respectively, but increases the largest sampled relative error to $3.86\%$. Overall, the performance of this re-optimisation effort is not satisfactory. 

\subsection{Wierling's continued fractions approximation}
\label{sec:wierling}

A closely related but distinct method of approximating the dielectric functions is the T-fraction approach of Wierling \cite{Wierling_PhysPlasmas_2009}. This is essentially a continued fraction generated from the known small- and large-argument expansions. The construction is of direct relevance here because T-fractions are closely connected to two-point Pad\'e approximants, but allow the two asymptotic series to be incorporated in a compact recursive representation.

Wierling writes the dynamic dielectric response function in terms of \eq{eq:H_def_3} but with the reduced versions of the arguments \eq{eq:Qkappa} considered by DAC

\begin{align}
	\label{eq:H_Wierling}
	\HR^{\rm Wier}(\chi;\eta)
	\equiv &\,
	\frac{h(x;\Theta)}{\sqrt{\pi\Theta}\,x\,\Fcal_{-1/2}(\eta)}
	\,,
	\\
	h(x;\Theta)
	= &\,
	\int_{0}^{\infty}
	\dif y\,
	\frac{y}{1 + e^{y^{2}/\Theta - \eta}}
	\ln\abs{\frac{x + y}{x - y}}
	\,,
\end{align}
which is identical to \eq{eq:H_def_3} once $q = \sqrt{\Theta}\varkappa$ is used. Note that the function $h(q,\eta)$ is used instead of Wierling's notation to avoid conflict with \eq{eq:G_def}.

For the static screening problem of interest, the continued fraction utilised by Wierling is written as
\begin{equation}
	\label{eq:wcf}
	h(x;\Theta)
	=
	\mathfrak{Re}\left\{
	\frac{\mu_0}{
		x-i b_1(\Theta)+
		\displaystyle\mathop{\mathrm K}_{n=2}^{\infty}
		\frac{i a_n(\Theta)x}{x-i b_n(\Theta)}
	}
	\right\}
	\,,
\end{equation}
with $\mu_0=2/3$, whose partial numerators and denominators are generated by a modified quotient-difference algorithm from the two asymptotic series. Wierling tabulates a set of coefficients in the case of $n=8$ at $\Theta \in [0,0.1,0.2,0.3,0.5,0.8,1,1.5,2.0]$ \cite{Wierling_PhysPlasmas_2009}. Importantly, there is no global fit to the $\Theta$ dependence published for the chosen implementation. The algorithm presented is of course general, and new coefficients and interpolations between them could be generated.

%% file: sections/main_body/05_new_model.tex
\section{Two-region approximation in compactified coordinates}

Neither of the preceding approximations provides the accuracy required for both static and dynamic applications across the classical, warm dense and strongly degenerate regimes. The principal difficulty is to reproduce the progressively sharper Kohn feature in $\HR$ and, especially, in its scaled derivative $\Ical=-\chi\partial\HR/\partial\chi$, without introducing poles or poorly conditioned coefficient functions. Since the real part of the dynamic dielectric response also depends sensitively on $\Ical$, particularly when the frequency-shifted arguments differ by small amounts, a two-regime representation is adopted. Each regime can then be described by an analytic form adapted to its dominant physical features.

The transition is chosen near the maximum rate of change of the Kohn-feature width with degeneracy, measured in the compactified coordinate $v=\mathcal{F}_{-1/2}(\eta)/\mathcal{F}_{1/2}(\eta)$. This occurs at approximately $v_{0}=0.2$, which is used as the boundary between the rational and composite-polynomial representations.

\subsection{Low-degeneracy regime: $v\geq v_{0}$ -- $6/7$ rational approximation}

For $v_{0}\leq v\leq v_{\rm class}$, where $v_{\rm class}=0.9991252223769$, the Kohn feature remains sufficiently broad to approximate both $\HR$ and $\Ical$ using a common rational function. We compactify the wave-number dependence through
\begin{align}
    \label{eq:u_def}
    u
    = &\,
    \frac{2v\chi^{2}}{1+2v\chi^{2}}
    ~\leftrightarrow~
    \chi
    =
    \sqrt{\frac{u}{2v(1-u)}}
    \,.
\end{align}
The interval $\chi\in[0,\infty)$ is thereby mapped onto $u\in[0,1)$. Since $\HR\sim(2v\chi^2)^{-1}\sim1-u$ at large $\chi$, the leading large-wave-number behaviour can be removed explicitly. The Kohn point $\chi^2=\eta$ is mapped to $u_{\rm K}=2v\eta/(1+2v\eta)$, which approaches $3/4$ in the strongly degenerate limit.

The resulting approximation is
\begin{align}
    \label{eq:H_fit_new}
    \HR^{\rm New}(\chi;\eta)
    \equiv &\,
    (1-u)\frac{P_6(u;v)}{Q_7(u;v,\eta)}
    \, ,
    \\
    \label{eq:P_new}
    P_6(u;v)
    = &\,
    1+\sum_{j=1}^{6}\alpha_j(v)u^j
    \, ,
    \\
    \label{eq:Q_new_2}
    Q_7(u;v,\eta)
    = &\,
    \sum_{k=0}^{7}\zeta_k(v,\eta)
    {7\choose k}u^k(1-u)^{7-k}
    \nnn
    = &\,
    1+\sum_{j=1}^{7}\beta_j(v)u^j
    \, .
\end{align}
Writing $R_{6/7}=P_6/Q_7$, its negative scaled derivative is evaluated analytically as
\begin{align}
    \Ical(u;v,\eta)
    = &\,
    2u(1-u)
    \left[R_{6/7}-(1-u)\frac{\partial R_{6/7}}{\partial u}\right]
    \, .
\end{align}

The exact small- and large-wave-number expansions through cubic order, \eqs{eq:smallk}{eq:largek}, determine all eight Bernstein coefficients $\zeta_{0},\ldots,\zeta_{7}$ algebraically from the numerator (see \app{app:H67_endpoint_constraints}). Hence the denominator does not require independent fitted functions. Moreover, its large-wave-number endpoint obeys $Q_7(1)=\zeta_7=P_6(1)$, so we write
\begin{align}
    \label{eq:alpha_6}
    \lim_{u\to1}\frac{\HR(u;v,\eta)}{1-u}
    = &\,
    \frac{P_6(1;v)}{\zeta_7(v)}=1
    \,,
    \\
    \label{eq:alpha_6_2}
    P_6(1;v)
    = &\,
    \zeta_7(v)=\exp[\xi(v)]>0
    \, .
\end{align}
This endpoint condition prevents the small positive value of $P_6(1; v)$ from having to be obtained by cancellation of the much larger alternating power-basis coefficients. The conversion to ordinary power coefficients, suitable for Horner evaluation, is
\begin{align}
    \beta_j(v)
    = &\,
    {7\choose j}
    \sum_{k=0}^{j}(-1)^{j-k}{j\choose k}\zeta_k(v,\eta)
    \, .
\end{align}

To exploit both endpoint values, $P_6(0; v)=1$ and $P_6(1; v)=\zeta_{7}(v)$, the numerator is written in the form
\begin{align}
    \label{eq:P_endpoint_anchored}
    P_6(u;v)
    = &\,
    1 + \rnd{\zeta_{7}(v) - 1}u
    \nnn &
    +
    u (1 - u) 
    \sum_{j=0}^{4} 
    s_j(v) u^j
    \, ,
    \\
    s_0(v)
    = &\,
    \alpha_1(v) - \rnd{\zeta_{7}(v) - 1}
    \, .
\end{align}
Given the exponential parametrisation $\zeta_{7}(v)=e^{\xi(v)}$, the numerically stabilised function $\mathrm{expm1}\rnd{\xi(v)}$ is used to evaluate $\zeta_{7}(v)-1$. The exact long-wavelength behaviour of $\HR$, which decreases from unity as $u\to0$, requires $\alpha_{1}-\beta_{1}-1<0$. Although this condition does not independently determine the sign of $\alpha_{1}$, it motivates imposing the additional restriction that the numerator also decreases initially. We therefore adopt $\alpha_{1}(v)=-e^{\gamma(v)}$, ensuring that $P_{6}'(0;v)<0$ whilst preserving the exact asymptotic constraints through the denominator coefficients. The signs of the remaining numerator coefficients are not independently constrained.

The conventional power-basis coefficients are then obtained from the relationships
\begin{align}
\label{eq:sj_def}
    \alpha_2(v) 
    = &\,
    s_1(v) - s_0(v)
    \,,
    \nnn
    \alpha_3(v) 
    = &\,
    s_2(v) - s_1(v)
    \,,
    \nnn
    \alpha_4(v) 
    = &\,
    s_3(v) - s_2(v)
    \,,
    \nnn
    \alpha_5(v) 
    = &\,
    s_4(v) - s_3(v)
    \,,
    \nnn
    \alpha_6(v) 
    = &\,
    -s_4(v)
    \,.
\end{align}

For each thermodynamic state, the fitted parameters are determined by minimising the maximum of the errors in the real boundary function $\HR$ and its real scaled derivative $\Ical$, subject to $Q_7(u)>0$. Specifically, defining $\cvec(v)=(s_1,s_2,s_3,s_4,\xi)^{\rm T}$ once $\alpha_1$ is prescribed, the error measures are
\begin{align}
    \label{eq:error_condition}
    \Ecal_{\HR}(v)
    = &\,
    \max_u\left|\frac{\Hcal_{\rm III}(u;v,\eta)-\Hcal_{\rm fit}(u;v,\eta)}{\Hcal_{\rm III}(u;v,\eta)}\right|
    \, ,
    \\
    \Ecal_{\Ical}(v)
    = &\,
    \frac{\displaystyle\max_u\left|\Ical_{\rm III}(u;v,\eta)-\Ical_{\rm fit}(u;v,\eta)\right|}{\displaystyle\max_{u'}\left|\Ical_{\rm III}(u';v,\eta)\right|}
    \, ,
\end{align}
with the joint objective $\max\{\Ecal_{\HR}(v),\Ecal_{\Ical}(v)\}\leq\Ecal_{\max}=0.1\%$ strictly enforced through the optimisation. Normalising the $\Ical$ error to its statewise maximum amplitude avoids singular relative errors near points where the function crosses zero.

Independent statewise optimisations may switch between nearly equivalent minima because the numerator coefficients are strongly correlated. A particularly strong mechanism for enforcing smooth, non-oscillatory and robust optimisation is a tightly branch-locked predictor--corrector continuation \cite{Rheinboldt1980, AllgowerGeorg1990}, from $v_{\rm class}$ down to $v_0$ using adaptive state stepping and local gradient and curvature information (see \app{app:branch_locked_fitting}). At each state, the prescribed exponential form of $\alpha_1(v)$ is imposed whilst the five remaining parameters are corrected along the solution branch. 

The statewise optima are fitted with Chebyshev polynomials in the following forms:
\begin{align}
    \label{eq:coefficients_fit}
    \alpha_1(v)
    = &\,
    -\exp\sqr{\sum_{n=0}^{4}
    A_n T_n(2\tilde{v}_{\rm class} - 1)}
    \,,
    \\
    s_j(v)
    = &\,
    \sum_{n=0}^{9} 
    B_{jn} T_n(2\tilde{v}_{\rm class} - 1),
    \qquad j=1,\ldots,4
    \,,
    \\
    \zeta_{7}(v)
    = &\,
    \exp\sqr{\sum_{n=0}^{9}
    C_{n} T_n(2\tilde{v}_{\rm class} - 1)}
    \,,
    \\
    \tilde{v}_{\rm class}
    = &\,
    \frac{v - v_0}{v_{\rm class} - v_0}
    \,.
\end{align}
Here $T_n(z)$ denotes a Chebyshev polynomial of the first kind, which provides a useful basis function for expansion of the solutions in the affine coordinate $\tilde{v}_{\rm class} \in [0, 1]$. The optimised coefficients $A_{n}$, $B_{jn}$ and $C_{n}$ are fitted very well, and are listed in \tab{tab:global_cheb55_coefficients}.

The resulting smooth statewise data were used to initialise a joint response-level optimisation of the global Chebyshev coefficients, with an additional constraint restricting the maximum departure of $\xi(v)$ from the branch-locked values to $0.10$. Positivity of $Q_7$ was imposed during fitting and checked numerically by Bernstein subdivision.

\begin{figure}[!t]
    \centering
    \includegraphics[width=0.49\textwidth]
    {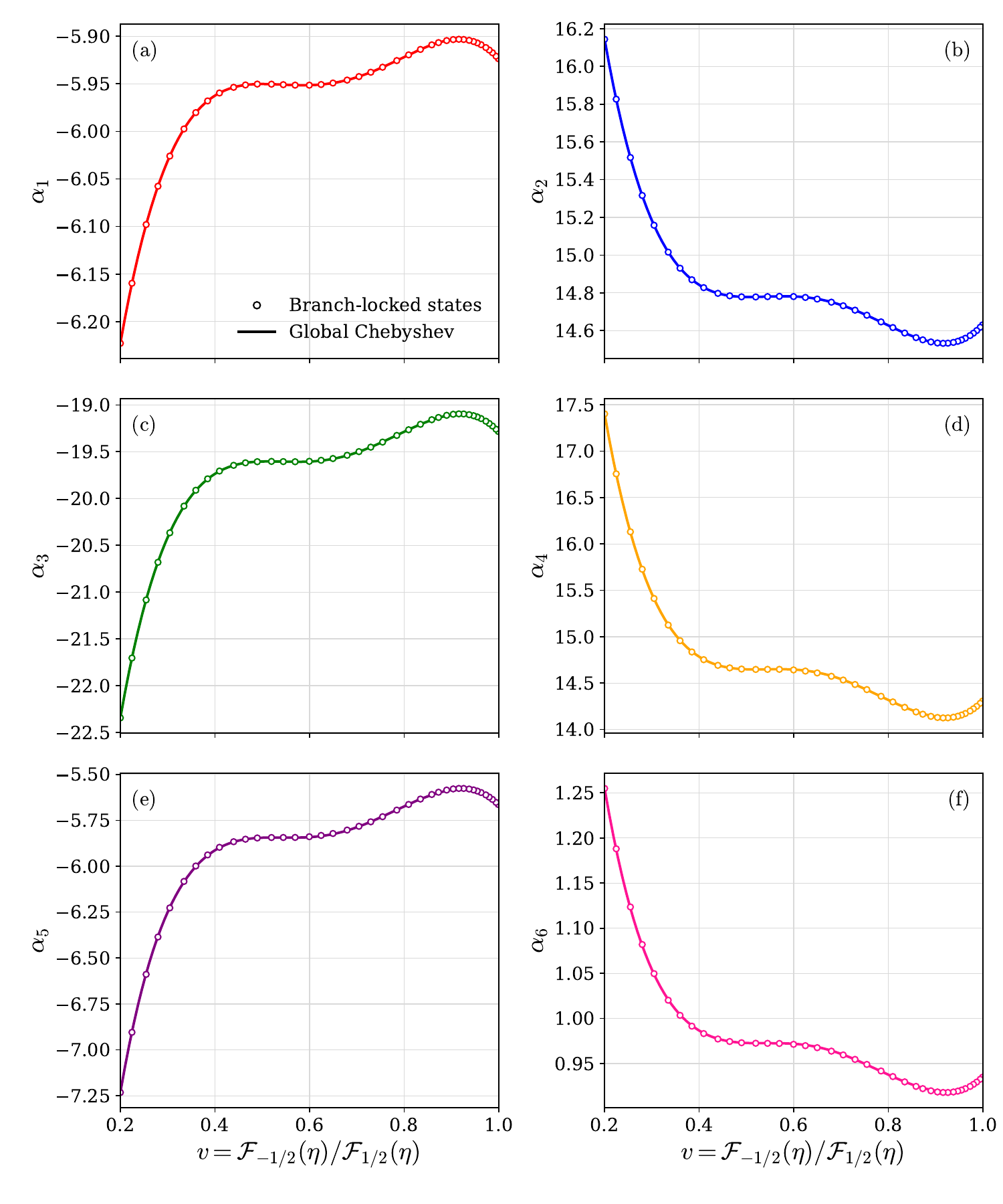}
    \caption{
        Power-basis numerator coefficients $\alpha_1,\ldots,\alpha_6$ of the endpoint-anchored $6/7$ rational approximation versus $v$. Open circles denote representative branch-locked statewise solutions, and solid curves show the single-interval Chebyshev parametrisation. The coefficients $\alpha_2,\ldots,\alpha_6$ are reconstructed from $\{s_1,\ldots,s_4,\xi\}$ by \eq{eq:sj_def}.
    }    \label{fig:low_degeneracy_regime_coefficients_fit}
\end{figure}

\begin{figure}[!t]
    \centering
    \includegraphics[width=0.49\textwidth]
    {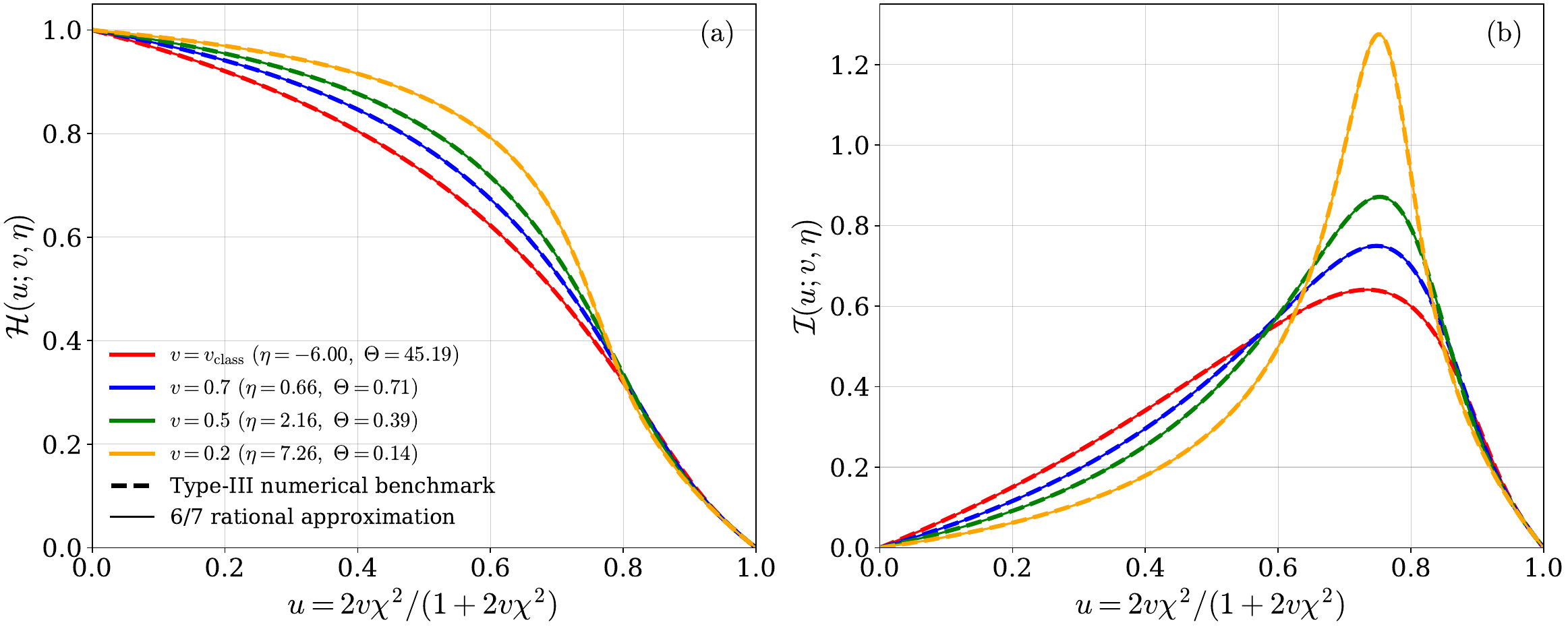}
    \caption{
        Comparison of the Type-III numerical benchmark (dashed curves) and the $6/7$ rational approximation (solid curves) for $\HR(u;v,\eta)$ (a) and $\Ical(u;v,\eta)$ (b) at $v=v_{\rm class}\approx1$ (red), $v=0.7$ (blue), $v=0.5$ (green) and $v=0.2$ (orange).
    }
    \label{fig:H_and_I_v_evolution_fit_vs_numerical}
    \vspace{0.7cm}
    \centering
    \includegraphics[width=0.49\textwidth]
    {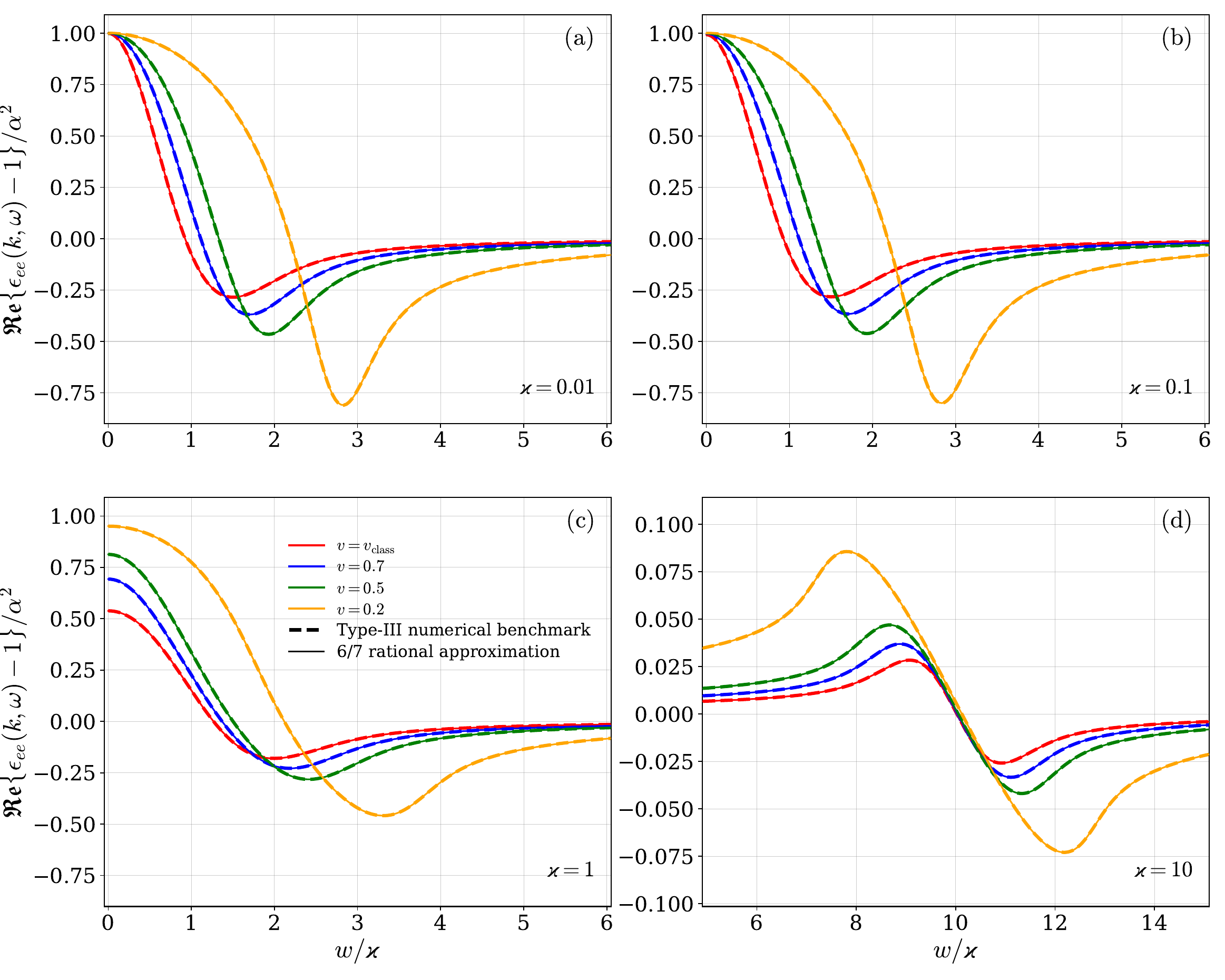}
    \caption{
        As in \fig{fig:H_and_I_v_evolution_fit_vs_numerical}, but for $\operatorname{Re}\{\epsilon_{ee}(k,\omega)-1\}/\alpha^2$ versus the dimensionless frequency shift $w/\varkappa$. Panels (a)--(d) correspond to $\varkappa=0.01$, $0.1$, $1$ and $10$, respectively.
    }
    \label{fig:Re_eps_v_evolution_fit_vs_numerical}
\end{figure}

The close agreement between the statewise data and the compact coefficient representation is shown in \fig{fig:low_degeneracy_regime_coefficients_fit}. A common-grid validation against the Type-III numerical benchmark, covering 301 degeneracy states and 1001 logarithmically spaced wave numbers $0.1\leq\chi\leq100$ per state, gives maximum relative errors of $0.08856\%$ in $\HR$ and $0.09387\%$ in the peak-normalised $\Ical$. On the same grid, the previously accepted 72-constant B-spline model gives $0.09116\%$ and $0.09940\%$, respectively. Thus the new 55-constant representation slightly improves the worst-case errors while reducing the number of fitted constants by $23.6\%$ and eliminating internal polynomial joins. An additional numerical denominator-positivity audit at 1001 states, including Bernstein-subdivision checks, found no non-positive values over the sampled domain; this is a numerical verification rather than a formal positivity proof for all continuous states.

The reduction in parametrisation complexity has little effect on evaluation time. With the same compiled response kernel and model-specific state preparation, the median times are $9.844\,\mu\mathrm{s}$ and $9.915\,\mu\mathrm{s}$ per 1001-wave-number state for the new Chebyshev and previous B-spline forms, respectively. The small difference (approximately $0.7\%$) is not significant relative to normal benchmarking variability. Both remain computationally inexpensive compared with direct numerical quadrature.

The results illustrate the evolution of the broad, low-amplitude feature in $\Ical$ near the classical limit into the narrower, higher-amplitude structure encountered on approaching $v_0$. The same rational functional form reproduces the smooth portions of $\HR$ throughout this interval. The representative dynamic-response comparisons in \fig{fig:Re_eps_v_evolution_fit_vs_numerical} illustrate the importance of jointly controlling $\HR$ and $\Ical$: the derivative constrains the response to small frequency-shift differences, whereas the correct asymptotic behaviour fixes its limiting values. The figures illustrate the response features targeted by the rational construction, while the static accuracy of the revised coefficient parametrisation is established independently by the numerical validation described above.

\subsection{High-degeneracy regime: $v < v_{0}$ - Composite polynomial approximation}

In the high-degeneracy regime the rapidly sharpening Kohn feature in $\Ical$ requires an approximation that explicitly resolves its structure. Although $\HR$ can be fitted reasonably well by a single smooth function, achieving sufficient accuracy in its derivative for the small-$\varkappa$ dynamic response is substantially more demanding. Rather than attempting to describe both $\HR$ and $\Ical$ simultaneously, it is easier to address the scaled derivative directly and integrate via
\begin{align}
	\label{eq:small_v_I_H_relationship}
	\HR(\chi;\eta)
	= &\,
	\int_{\chi}^{\infty}
	\frac{\dif \chi'}{\chi'}
	\Ical(\chi';\eta)
	\,,
\end{align}
to achieve a consistent fit to $\HR$. A final form that has an exact antiderivative without using special functions is therefore highly desirable. The basis of the model in this regime is given by the decomposition 
\begin{align}
	\label{eq:Hcal_decomp}
	\HR(\chi;\eta)
	-
	\frac{g(\chi_{0})}{2\chi_{0}}
	= &\,
	\Delta_{T}\HR(u;v,\eta)
	\nnn
	= &\,
	\HR^{\Kcal}(\chi, u;v,\eta)
	+
	\HR^{\Rcal}(u;v,\eta)
	\nnn
	= &\,
	\frac{1}{2}
	\int_{u}^{1}
	\frac{\dif u'}{u'(1 - u')}
	\left[
		\Ical^{\Kcal}(\chi_{0}',u';v,\eta)
		\right. \nnn & \left.
		+
		\Ical^{\Rcal}(u';v,\eta) 
	\right]
	\,.
\end{align}
The form of the integrals results from the change of variable from $\chi$ to the compactified coordinate $u$. Thus, the aim is to approximate the thermal correction to the zero-temperature Lindhard result. 

\begin{figure}[!t]
	\centering
	\includegraphics[width=0.44\textwidth]
	{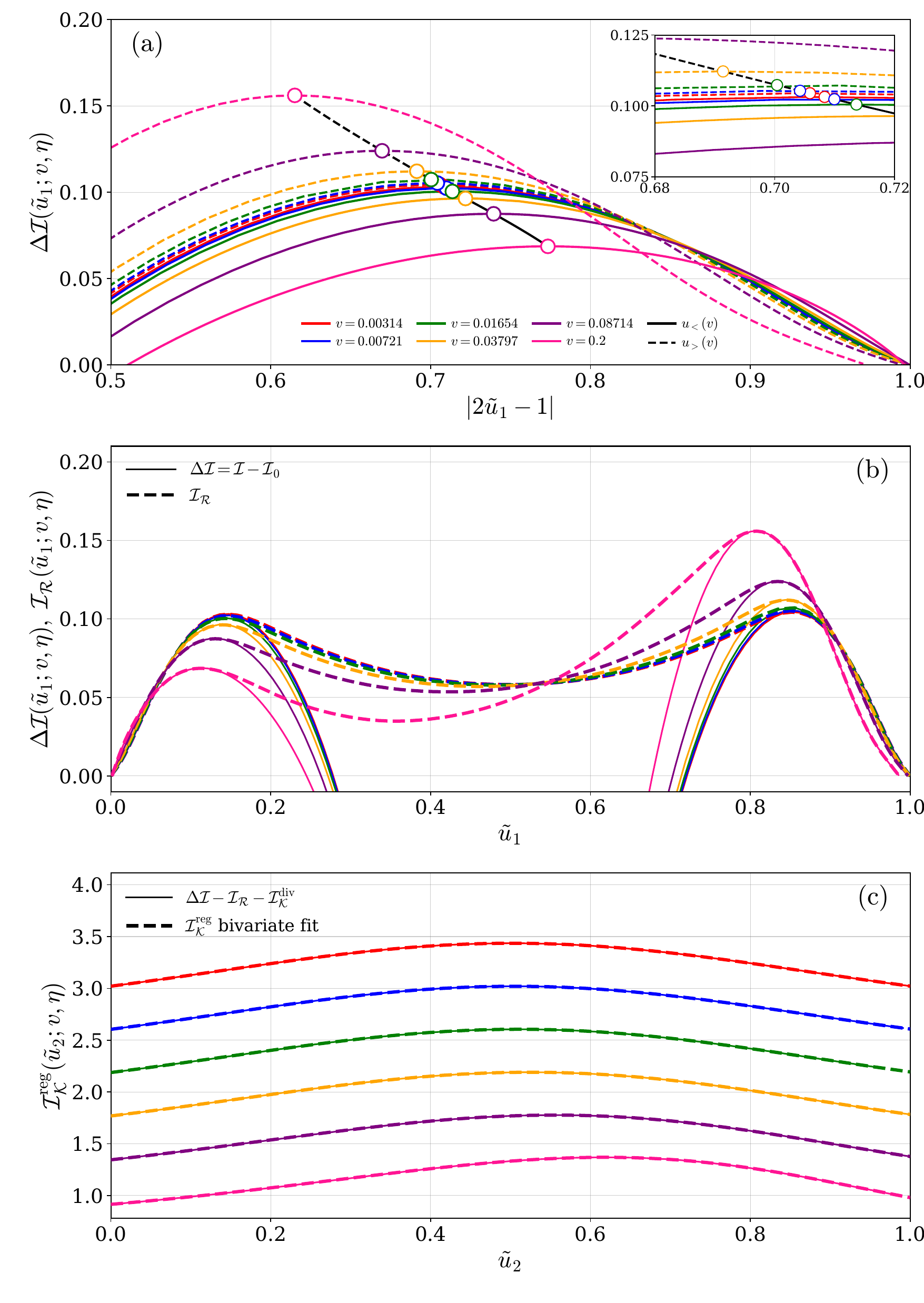}
	\caption{
		Difference of the scaled derivative at finite and zero temperatures $\Delta\Ical = \Ical - \Ical_{0}$ (a) as a function of the absolute scaled coordinate $\abs{2\tilde{u}_{1} - 1}$ for values of $v$ spanning the high-degeneracy regime $v_{\rm degen} < v < v_{0}$. The solid curves correspond to the left-hand branch and the dashed curves to the right-hand branch. The fits to the stationary points $u_{\gtrless}$ \eqref{eq:u_gtrless} are plotted using black curves with matching styles. Panel (b) shows $\Delta\Ical$ in addition to the smooth residual term $\Ical^{\Rcal}$, constructed via the procedure described in the text that connects the shoulders either side of the Kohn regime (dashed curves). The regular part of the Kohn contribution $\Ical^{\Kcal}_{\rm reg}$ is shown in Panel (c) along with its bivariate polynomial fit.
	}
	\label{fig:I_approximation_construction_stages}
\end{figure}

It is useful to partition the correction term $\Delta_{T}\Ical = -\chi\Delta_{T}\HR'$ into a contribution that captures the divergent Kohn feature $\Ical^{\Kcal}$ and some smooth residual $\Ical^{\Rcal}$. A pragmatic approach identifies the edges of the region encompassing the divergent feature (see the markers at the points of zero gradient either side of the central negative-valued peak in \fig{fig:I_approximation_construction_stages}(a)) and tracks their evolution with degeneracy. This can be parametrised via
\begin{align}
	\label{eq:u_gtrless}
	u_{\gtrless}(v)
	= &\,
	\frac{3}{4} + v \sum_{n = 0}^{3} D_{n}^{\gtrless}v^{n}
	\,.
\end{align}
The width of this region is then
\begin{align}
	\label{eq:Delta_u_K}
	\Delta u_{\rm K}(v) 
	= &\,
	u_{>}(v) - u_{<}(v)
	=
	v\sum_{n = 0}^{3}\rnd{D_{n}^{>} - D_{n}^{<}}v^{n}
	\,.
\end{align}
The coefficients $D_{n}^{\gtrless}$ are tabulated in \tab{tab:new_kohn_boundary_coefficients}.

In the region containing the Kohn feature the contribution is broken into divergent and regular parts:
\begin{align}
	\label{eq:Kcal}
	\Ical^{\Kcal}(\chi_{0},u;v,\eta)
	= &\,
	\rectpulse(\chi_{0}; \chi_{0<}, \chi_{0>})
	\Ical^{\Kcal}_{\rm div}(\chi_{0}) 
	\nnn &
	+
	\rectpulse(u; u_{<}, u_{>})
	\Ical^{\Kcal}_{\rm reg}(u;v,\eta)
	\,,
\end{align}
where the symbol $\rectpulse(x;a,b) \equiv \Theta(x - a)\Theta(b - x)$ is the square window filter, which is unity for $a \leq x \leq b$ and zero otherwise, and $\chi_{0\gtrless} = \chi_{\gtrless}/\sqrt{\eta} = \sqrt{u_{\gtrless}/2v\eta(1 - u_{\gtrless})}$.

Identifying the logarithmically divergent part as
\begin{align}
	\Ical^{\Kcal}_{\rm div}(\chi_{0})
	= &\,
	\frac{1 + \chi_{0}^{2}}{4\chi_{0}}\ln\abs{1 - \chi_{0}}
	\,,
\end{align} 
then the corresponding integrated term, which is itself not divergent, is
\begin{align}
	{\HR^{\Kcal}}_{\rm div}(\chi_{0}) 
	= &\,
	\Theta(\chi_{0>} - \max(\chi_{<0},\chi_{0}))
	\nnn &
	\times
	\sqr{
		\Xi(\chi_{0>}) 
		- 
		\Xi(\max(\chi_{<0},\chi_{0}))
	}
	\,,
	\\
	\Xi(x)
	= &\,
	\frac{1}{4}
	\sqr{
		\frac{x^{2} - 1}{x}\ln\abs{1 - x}
		-\ln x - x
	} 
	\,.
\end{align}
Note that at the singular point $\chi_{0} = 1$ the function $\Xi(1) = -1/4$. This leaves the two functions $\Ical^{\Rcal}$ and $\Ical^{\Kcal}_{\rm reg}$ to fit. 

The smooth remainder term $\Ical^{\Rcal}$ is constructed first in two distinct stages. First, the thermal correction term $\Delta\Ical$ is fitted to high accuracy using Chebyshev polynomials of the first kind $T_{n}(z)$ for $u < u_{<}$ and $u > u_{>}$, subject to the conditions that their gradients disappear at $u = u_{<}$ and $u = u_{>}$. Within the Kohn region, the polynomials are joined with a fifth order Hermite interpolant in order to match the values, derivatives and curvatures, resulting in a $C^2$ continuous function when all three regions are piecewise joined (\fig{fig:I_approximation_construction_stages}(b)). The central dip arises as a consequence of the negative curvatures at the zero-gradient points. The second stage then fits the joined three-region result with the prescribed form
\begin{align}
	\label{eq:Ical_Rcal}
	\Ical^{\Rcal}(u;v,\eta)	
	= &\,
	u(1 - u)
	\Qcal^{\Rcal}(u;v,\eta)	
	\,,
\end{align}
motivated by the fact that the prefactor $u(1-u)$ cancels with the Jacobian term of the integral in \eq{eq:Hcal_decomp}. The result for the polynomial part is
\begin{align}
	\label{eq:Qcal_R}
	\Qcal^{\Rcal}(u;v,\eta)	
	= &\,
	\frac{64}{3}
	\tilde{u}_{1}(1 - \tilde{u}_{1})
	\sum_{n = 0}^{N_{\Rcal}}
	E_{n}
	T_{n}(2\tilde{u}_{1} - 1)
	\,,
	\\
	E_{n}
	= &\,
	\sum_{m = 0}^{M_{\Rcal}}
	E_{nm}T_{m}(2\tilde{v}_{\rm degen} - 1) 
	\,.
\end{align}
In \eq{eq:Qcal_R}, the coordinate transforms 
\begin{align}
	\tilde{u}_{1}
	= &\,
	\frac{1}{2}
	\left[
	1 + \frac{u - u_{\rm K}}{\sqrt{\rnd{u - u_{\rm K}}^{2} + (\Delta u_{\rm K}/2)^{2}}}
	\right]
	\,,
	\\
	\tilde{v}_{\rm degen}
	= &\,
	\frac{v - v_{\rm degen}}{v_{0} - v_{\rm degen}}
	\,,
\end{align}
enable mapping each component of the fits to the interval $z \in [-1, 1]$, as required by the Chebyshev polynomials. The first in particular, $\tilde{u}_{1}$, helps to focus the fit to the shrinking width of the Kohn region in the limit $v \to 0$. The antiderivative of \eq{eq:Qcal_R} is analytic, with the result
\begin{align}
	\HR^{\Rcal}(u;v,\eta)
	= &\,
	\frac{4}{3}\Delta u_{\rm K}
	\sqr{
		\Phi(\tilde{u}_{1}(u); v, \eta)
	}_{1}^{u}
	\,,
	\\
	\Phi(\tilde{u}_{1};v,\eta)
	= &\,
	E_{0}\arccos(2\tilde{u}_{1} - 1) + 2\sqrt{\tilde{u}_{1}(1 - \tilde{u}_{1})}
	\nnn & 
	\times
	\sum_{n = 1}^{N_{\Rcal}}\frac{E_{n}}{n}U_{n - 1}(2\tilde{u}_{1} - 1)
	\,,
\end{align}
in which $U_{n}(z)$ are the Chebyshev polynomials of the second kind.

The final component required for the high-degeneracy model is the fit for the regular part in the Kohn region $\Ical^{\Kcal}$. Similarly to \eq{eq:Ical_Rcal} the form is asserted to be
\begin{align}
	\Ical^{\Kcal}_{\rm reg}(u;v,\eta)	
	= &\,
	u(1 - u)
	\Qcal^{\Kcal}_{\rm reg}(u;v,\eta)	
	\,.
\end{align}
However, an additional constraint must also be respected by the polynomial part in this case; the total contribution $\Ical^{\Kcal}$ must reach its target value of zero at both edges of the Kohn region, whilst also maintaining the gradient and curvature of the smooth residual, $\Ical^{\Rcal}$, i.e., 
\begin{align}
	\left.\frac{\partial^{j}}{\partial u^{j}}
	[\Ical^{\Kcal}_{\rm div}(\chi_{0}) + u(1 - u)
	\Qcal^{\Kcal}_{\rm reg}(u;v,\eta)]\right|_{u = u_{\gtrless}}
	= 0
	\,,
\end{align}
for $j = 0,1,2$. The polynomial part is therefore split into a fifth order Hermite interpolant $\text{Herm}[\Ical^{\Kcal}_{\rm div}]$ that automatically fulfils the derivative requirements of the function $-\Ical^{\Kcal}_{\rm div}(\chi_{0}(u))/u(1-u)$ (without introducing more free parameters that require optimisation) and a second polynomial contribution. Importantly, the latter must be constructed specifically as to not disturb the values, derivatives and curvatures established by the interpolant part. The lowest order factorisation that produces the desired behaviour is
\begin{align}
	\label{eq:Qcal_K_reg}
	\Qcal^{\Kcal}_{\rm reg}(u;v,\eta)	
	= &\,
	\tilde{u}_{2}^{3}(1 - \tilde{u}_{2})^{3}
	\sum_{n = 0}^{N_{\Kcal}}
	F_{n}
	T_{n}(2\tilde{u}_{2} - 1)
	\nnn &
	+
	\mathrm{Herm}[\Ical^{\Kcal}_{\rm div}](\tilde{u}_{2};v,\eta)
	\,,
	\\
	F_{n}
	= &\,
	\sum_{m = 0}^{M_{\Kcal}}
	F_{nm}T_{m}(2\tilde{v} - 1) 
	\,,
	\\
	\label{eq:tilde_u_2}
	\tilde{u}_{2}
	= &\,
	\frac{u - u_{<}}{\Delta u_{\rm K}}
	\,.
\end{align}
Analytically integrating gives
\begin{align}
	\label{eq:H_high_degeneracy}
	{\HR^{\Kcal}}_{\rm reg}(u;v,\eta)
	= &\,
	\frac{1}{2}\Delta u_{\rm K}
	\sqr{\Psi(\tilde{u}_{2}(u);v,\eta)}_{u_{\min}}^{1}
	\,,
	\\
	\Psi(\tilde{u}_{2};v,\eta)
	= &\,
	\sum_{n = 0}^{5}
	\frac{h_{n}(v,\eta)}{n + 1}
	\tilde{u}_{2}^{n + 1}
	\nnn &
	+
	F_{0} \Jcal_{0}(\tilde{u}_{2})
	+
	\sum_{n = 1}^{N_{\Kcal}}
	F_{n} \Jcal_{n}(\tilde{u}_{2})
	\,,
	\\
	\Jcal_{0}(\tilde{u}_{2})
	= &\,
	\tilde{u}_{2}^{4}
	\sqr{
		\frac{1}{4}
		-
		\frac{3\tilde{u}_{2}}{5}
		+
		\frac{\tilde{u}_{2}^{2}}{2}
		-
		\frac{\tilde{u}_{2}^{3}}{7}
	}
	\,,
	\\
	\Jcal_{n \geq 1}(\tilde{u}_{2})
	= &\,
	\sum_{k = 0}^{n}(-1)^{n - k}
	\frac{4^{k}n}{n + k}{n + k \choose 2k}
	\tilde{u}_{2}^{k + 4}
	\nnn &
	\times
	\sqr{
		\frac{1}{k + 4}
		-
		\frac{3\tilde{u}_{2}}{k + 5}
		+
		\frac{3\tilde{u}_{2}^{2}}{k + 6}
		-
		\frac{\tilde{u}_{2}^{3}}{k + 7}
	}
	\,.
\end{align}
The value of $u$ in the transformed coordinate at the lower integration limit is $u_{\min} = \min[u_{>},\max(u_{<},u)]$. The coefficients of the Hermite interpolant are all specified in terms of the derivatives of $-\Ical^{\Kcal}_{\rm div}/u(1-u)$ (see \app{app:hermite_interpolant}).

The four integers specifying the orders of the fit, $(N_{\Rcal}, M_{\Rcal})$ and $(N_{\Kcal}, M_{\Kcal})$, are determined along with the matrices of coefficients $\underline{\Evec}$ and $\underline{\Fvec}$ using the same condition employed in \eq{eq:error_condition} for the low-degeneracy regime. Here, they are $(20, 3)$ and $(4, 4)$. The coefficients $E_{nm}^{(20,3)}$ and $F_{nm}^{(4,4)}$ are given in \taba{tab:new_R_background_a}{tab:new_kohn_core_coefficients}.

\begin{figure}[!t]
	\centering
	\includegraphics[width=0.48\textwidth]
	{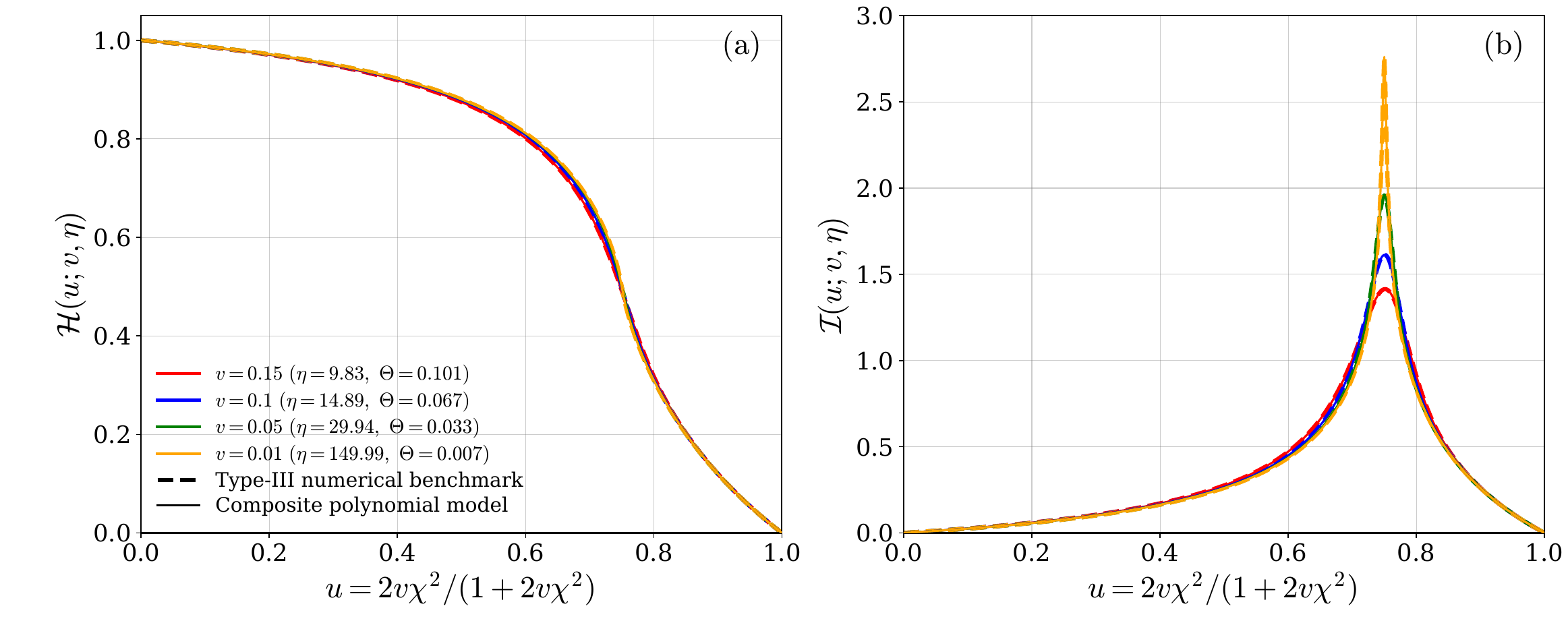}
	\caption{
		As in \fig{fig:H_and_I_v_evolution_fit_vs_numerical} but for states spanning the high-degeneracy regime: $v = 0.15$ (red); $v = 0.1$ (blue); $v = 0.05$ (green); and $v = 0.01$ (orange).
	}
	\label{fig:H_and_I_v_evolution_fit_vs_numerical_high_degen}
	%
	\vspace{0.7cm}
	\centering
	\includegraphics[width=0.48\textwidth]
	{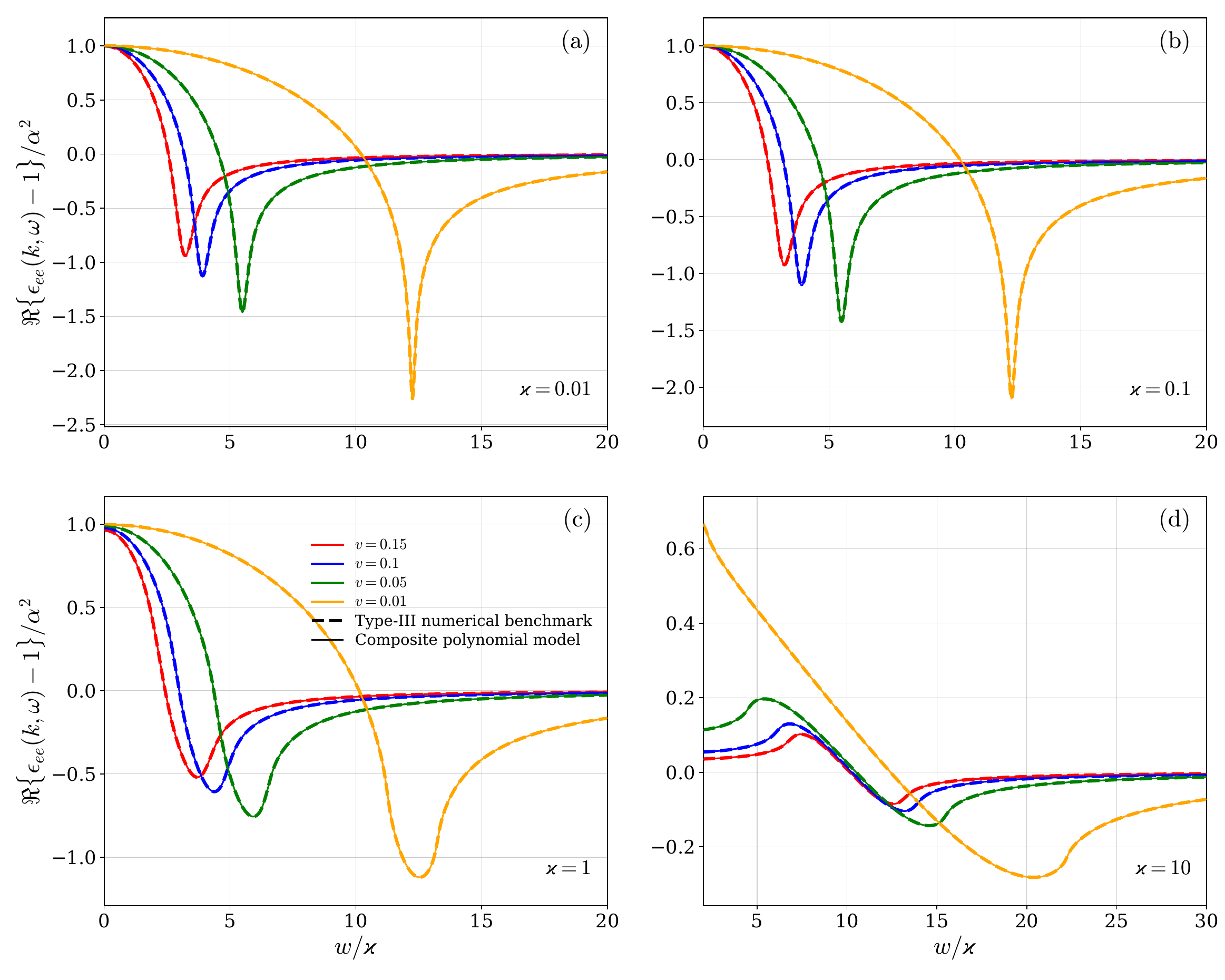}
	\caption{
		As in \fig{fig:Re_eps_v_evolution_fit_vs_numerical} but for states spanning the high-degeneracy regime: $v = 0.15$ (red); $v = 0.1$ (blue); $v = 0.05$ (green); and $v = 0.01$ (orange).
	}
	\label{fig:Re_eps_v_evolution_fit_vs_numerical_high_degen}
\end{figure}

The agreement between the composite approximation and the numerical benchmarks is excellent throughout the high-degeneracy regime (\fig{fig:H_and_I_v_evolution_fit_vs_numerical_high_degen}). The fitted $\HR$ curves are virtually indistinguishable from the numerical results across the full range of the compactified coordinate $u$, accurately reproducing both the smooth long-wavelength behaviour and the increasingly sharp variation associated with the Kohn feature. The corresponding agreement for $\Ical$ is particularly encouraging, since the narrowing and increasing amplitude of its characteristic peak is a substantially more demanding approximation problem. Even at $v=0.01$, where the feature becomes highly localised, the composite polynomial representation reproduces its position, amplitude and surrounding structure with high fidelity. This demonstrates the effectiveness of separating out the Kohn contribution.

The quality of the approximation is further demonstrated by the real part of the scaled dynamic dielectric function, $\mathfrak{Re}\{\epsilon_{ee}(k,\omega)-1\}/\alpha^2$ (\fig{fig:Re_eps_v_evolution_fit_vs_numerical_high_degen}). The agreement is maintained over the wave numbers considered, including the negative-response region where the pronounced feature develops. These features are more poorly modelled using the other fits mentioned in this work, especially for high degeneracy. The composite model captures the location and shape of this feature as well as the decay towards zero at large frequency shifts. Of particular note is the absence of appreciable numerical artefacts around the Kohn feature, despite the rapid variation of the underlying functions. This confirms that the accuracy achieved for the static response is successfully propagated into its dynamic counterpart.

Together with the complementary low-degeneracy rational approximation, the composite polynomial construction provides a unified description over the targeted range, from the classical regime to strong degeneracy. The new global-Chebyshev rational branch meets the $0.1\%$ criterion for both $\HR$ and $\Ical$ on the independent low-degeneracy validation grid. Since the high-degeneracy branch is unchanged, the maximum relative error in $\HR$ over the original full-domain benchmark remains approximately $0.109\%$, with the largest discrepancies localised near the handover between representations. The simplified coefficient parametrisation therefore improves the low-degeneracy model without changing the established accuracy of the composite high-degeneracy approximation.

%% file: sections/main_body/06_results_and_applications.tex
\section{Results and applications}

\subsection{Accuracy and efficiency}

\begin{figure}[!t]
	\centering
	\includegraphics[width=0.49\textwidth]
	{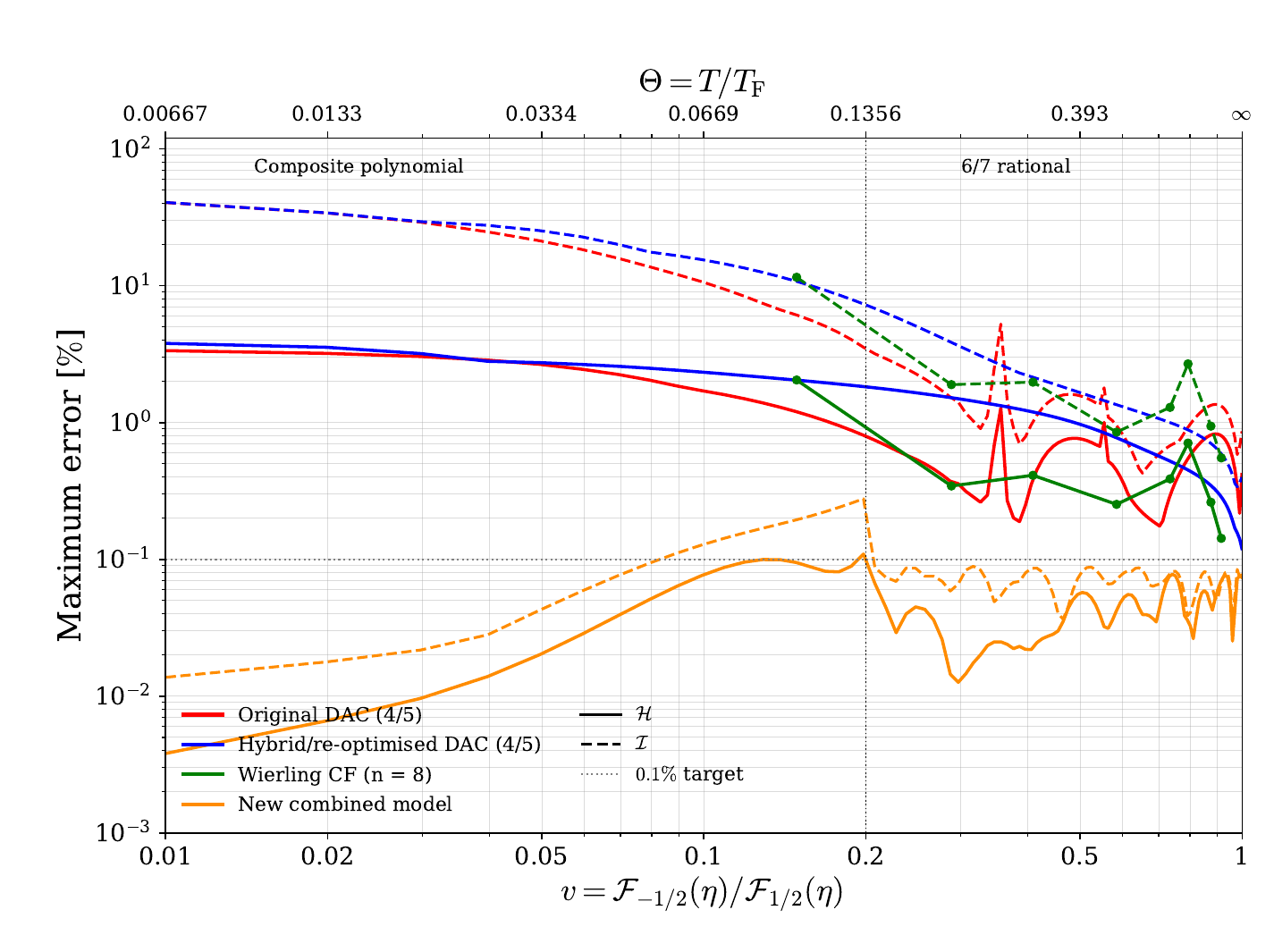}	
	\caption{
		Maximum relative error in $\HcalR$ (solid curves) and $\Ical$ (dashed curves) over a logarithmic grid of $\chi \in [0.1, 100]$ (1001 points) at the finite-temperature states tabulated by Wierling (filled circle markers). The continued fraction approach is competitive with, and often more accurate than, DAC under weak and moderate degeneracy, but its error rises again as the Kohn region sharpens. The new model is clearly the most accurate for all states, particularly when $v < v_{0}$.
	}
	\label{fig:werr}
\end{figure}

A comparison of the models considered in this work for the maximum relative error as a function of degeneracy is shown in \fig{fig:werr}. The points with markers coincide with those for which tabulated coefficients are given by Wierling to enable a direct comparison to the purely rational approximations and the new model. With respect to $\HcalR$, the latter is clearly the most globally accurate in terms of the $L_{\infty}$ errors, especially in the high-degeneracy regime where the other models struggle to capture the structure at the Kohn point. This is emphasised even more vividly when considering $\Ical$, since in this case the distributions' maximum errors are significantly elevated, reaching $40\%$ for the purely rational models at $v = 0.01$. In contrast, the new model is at just over $0.01\%$. Its worst case global errors are just over $0.1\%$ in $\HcalR$ and slightly less than $0.3\%$ in $\Ical$. Both of these points occur just after the transition point between degeneracy regimes, i.e., for $v \lesssim 0.2$.

\begin{figure*}[!t]
	\centering
	\includegraphics[width=0.93\textwidth]
	{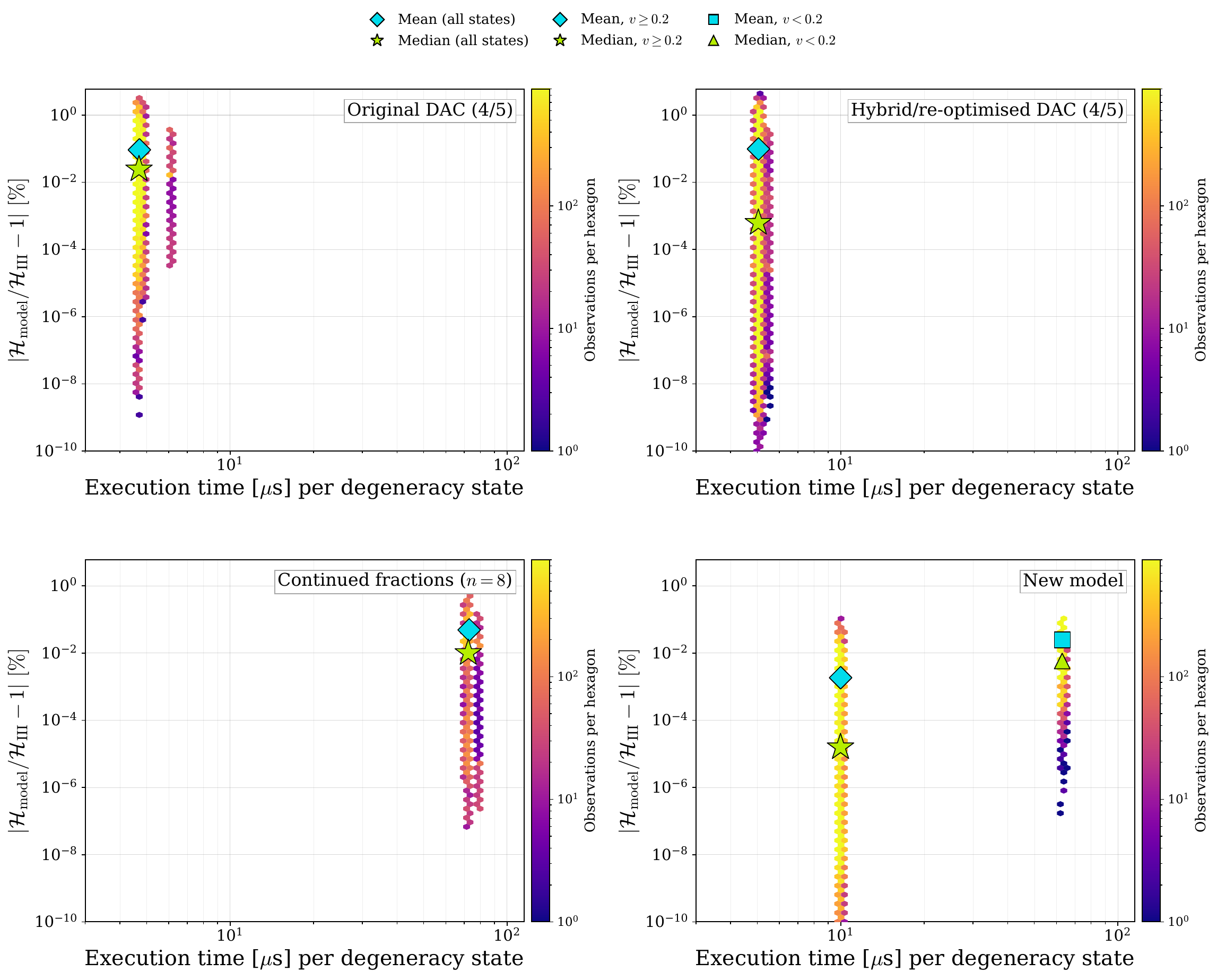}
	\caption{
		Relative absolute percentage error (compared to the Type III numerical benchmark) versus execution time per degeneracy state (as measured by $v$) for the four models considered in this work rendered as hexbin plots. For the three rational approximations $v \in [0.01, 1)$ with 101 values are shown, whilst only the eight non-zero $\Theta$ values provided by Wierling are used for the continued fraction approach. All models use 1001 values for $\chi \in [0.1, 100]$. The median and mean values are shown by lime green stars and cyan diamonds. For the new model, these show the performance in the low-degeneracy regime (faster cluster), whilst the cyan square and lime green triangle are the equivalent measures in the high-degeneracy regime (slower cluster).
	}
	\label{fig:error_vs_timing_model_comp}
\end{figure*}

Surprisingly, the hybrid/re-optimised approach performs poorly by all measures until $\Theta > 1$, despite being constructed using accurate special functions which matter most for intermediate degeneracies. This is a strong indication that the construction of the interpolated forms of the coefficients is of greater importance than improving the accuracy of the FDIs. DAC's original model appears to be quite accurate and of similar quality to Wierling's continued fraction method. 

A full audit of the accuracy and efficiency of all the models considered in this work is presented as hexbin plots in \fig{fig:error_vs_timing_model_comp}. All the models are evaluated and timed using the configuration corresponding to their most common application: once per state point (consumed via $v$, $\eta$ or $\Theta$) over an array of $\chi$ values ($\chi \in [0.1, 100]$). The DAC, hybrid/re-optimised, and new models all consider 101 values $v \in [0.01, 1)$, whilst the continued fraction method is restricted to the eight non-zero $\Theta$ points published by Wierling.

The data collected show the trade-off between accuracy and efficiency of the various models. The most algebraically simple (and subsequently fastest) are the purely rational models, which are approximately even in execution time, having mean/median results of $4.565\,\unit{\mu s}$/$4.567\,\unit{\mu s}$ for DAC and $4.871\,\unit{\mu s}$/$4.877\,\unit{\mu s}$ for the hybrid/re-optimised model. The new model is roughly twice as slow at low degeneracy at $9.841\,\unit{\mu s}$/$9.844\,\unit{\mu s}$, but far slower for high degeneracy at $62.694\,\unit{\mu s}$/$62.655\,\unit{\mu s}$. This is a direct consequence of the substantially larger number of arithmetic operations required for the composite polynomial approximation. However, this short fall is more than compensated for in its global accuracy, and even at its most expensive is faster than Wierling's continued fraction approach, which takes $81.211\,\unit{\mu s}$/$80.076\,\unit{\mu s}$. The comparative slowness of the latter is due to the requirement of evaluating the nested continued fractions themselves.

For comparison, the distribution of execution times for the Type III benchmark spans a range of tens to thousands of times slower, with median and mean execution times of $751.5\,\unit{\mu s}$ and $919.6\,\unit{\mu s}$, respectively. Reducing the integration tolerance to the same level as the target used for the new model ($0.1\%$) reduces these to $390.5\,\unit{\mu s}$ and $539.1\,\unit{\mu s}$; still orders of magnitude more numerically expensive.

\subsection{Dynamic structure factor and imaginary-time correlation function}

A stringent test of the present model is obtained by propagating the approximation to $\HcalR$ through the full dynamic dielectric response. The corresponding electron--electron dynamic structure factor (DSF) is central to X-ray Thomson scattering (XRTS) and related descriptions of the electronic density response of warm dense matter \cite{Crowley_HEDP_2014b,Dornheim_PhysPlasmas_2023a, Bellenbaum_PhysPlasmas_2026}. In the present variables, the reduced DSF
\begin{align}
	\label{eq:dsf_def}
	\tilde{S}_{ee}(\varkappa,w;\alpha)
	\equiv &\,
	\frac{\pi n_e e^{2}}{\epsilon_{0} \hbar k^{2}}S_{ee}(k,\omega)
	\nnn
	= &\,
	\frac{L(\varkappa,w;\alpha)}
	{1-e^{-4w}}
	\,,
	\\
	\label{eq:L_def}
	L(\varkappa,w;\alpha)
	= &\,
	\mathfrak{Im}
	\left\{
	-\frac{1}{\epsilon_{ee}(\varkappa,w;\alpha)}
	\right\}
	\,,
\end{align}
is introduced. The collisionless dynamic dielectric function samples the real boundary $\HcalR$ at the displaced arguments $w/\varkappa\pm\varkappa$, while its imaginary part is evaluated exactly by \eq{eq:H_exact_imag}. Consequently, even small approximation errors in $\HcalR$ can be amplified by the nonlinear dependence of the loss function on the inverse dielectric response, particularly near weakly damped collective excitations.

\begin{figure}[!t]
	\centering
	\includegraphics[width=\columnwidth]
	{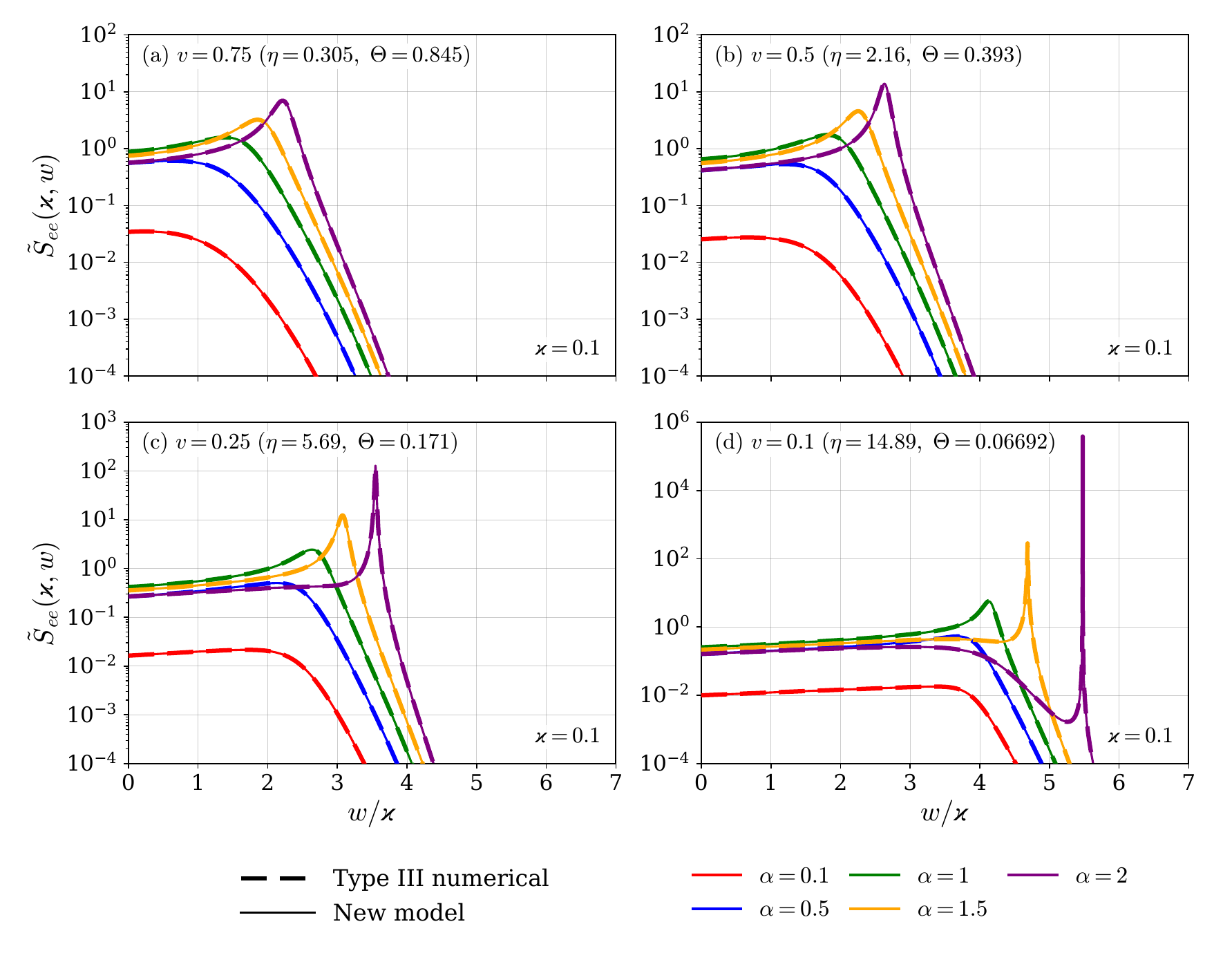}
	\caption{
		Reduced electron--electron dynamic structure factor, $\tilde{S}_{ee}(\varkappa,w)$, as a function of $w/\varkappa$ for $\varkappa=0.1$ and Salpeter collectivity parameters of $\alpha\in\{0.1,0.5,1,1.5,2\}$. The solid curves show the new model and dashed curves the numerical benchmark. The four panels correspond to: (a) $v=0.75$; (b) $v=0.5$; (c) $v=0.25$; and (d) $v=0.1$.
	}
	\label{fig:dsf_comparison_kap=0.1}
\end{figure}

\begin{figure}[!t]
	\centering
	\includegraphics[width=\columnwidth]
	{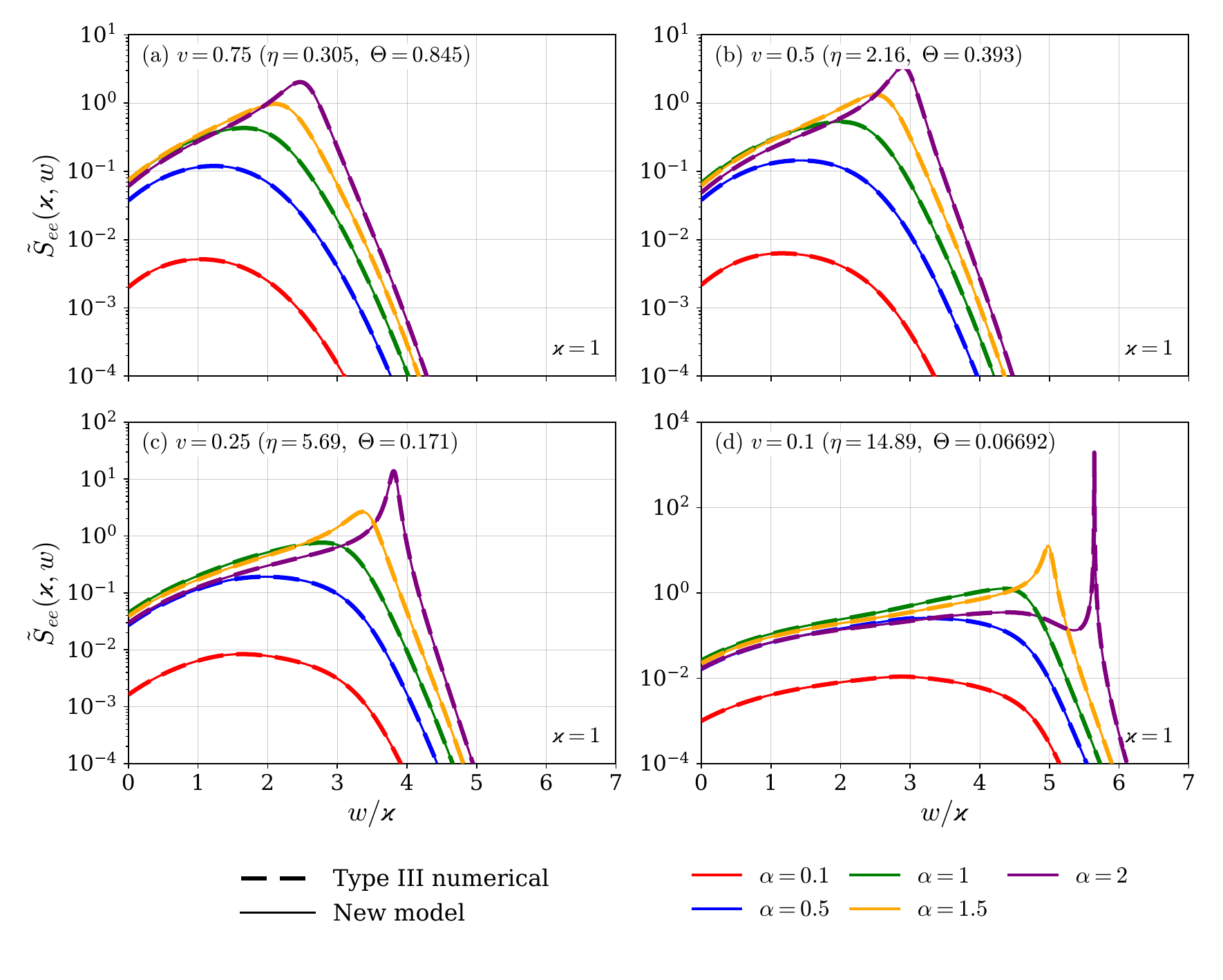}
	\caption{
		As in \fig{fig:dsf_comparison_kap=0.1}, but for
		$\varkappa=1$.
	}
	\label{fig:dsf_comparison_kap=1}
\end{figure}

\figa{fig:dsf_comparison_kap=0.1}{fig:dsf_comparison_kap=1} demonstrate the excellent agreement between the new model and the numerical benchmark across both degeneracy regimes. The agreement extends beyond the smooth continuum response to the positions, widths and amplitudes of collective plasmon resonances. In particular, the introduction of the high-degeneracy composite polynomial representation has substantially improved the description of the narrow spectral features that previously challenged global fits.

Across the 40 cases considered, 33/40 reproduce the numerical DSF peak amplitude to better than $0.1\%$, while 39/40 remain within $1\%$. The largest peak-height discrepancy is just $1.04\%$, which occurs at $(\varkappa,v,\alpha)=(0.1,0.1,2)$. For $\varkappa=1$, the largest peak-height error is only $0.49\%$. The remaining discrepancies are concentrated around very narrow, weakly damped plasmons. In this regime, the loss function is particularly sensitive to small shifts in the frequency at which $\mathfrak{Re}\{\epsilon_{ee}(k,\omega)\}$ vanishes. Accordingly, the pointwise difference between two narrow spectral peaks can be appreciable even when their maximum amplitudes and resonance frequencies agree closely. Peak-height errors alone therefore do not fully characterise the local spectral discrepancy.

\begin{figure}[!t]
	\centering
	\includegraphics[width=\columnwidth]
	{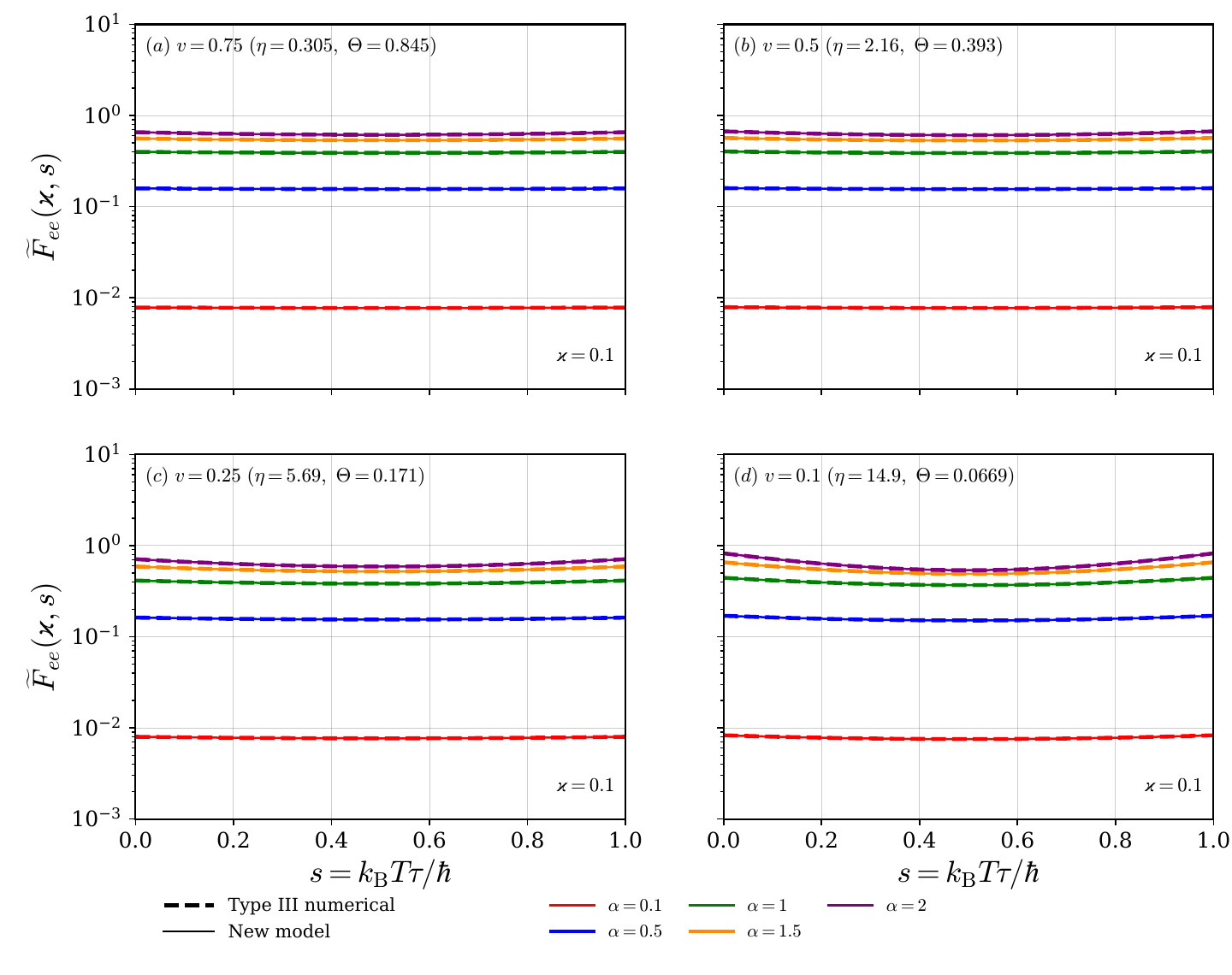}
	\caption{
		As in \fig{fig:dsf_comparison_kap=0.1}, but showing the ITCF instead of the DSF.
	}
	\label{fig:itcf_comparison_kap=0.1}
\end{figure}

\begin{figure}[!t]
	\centering
	\includegraphics[width=\columnwidth]
	{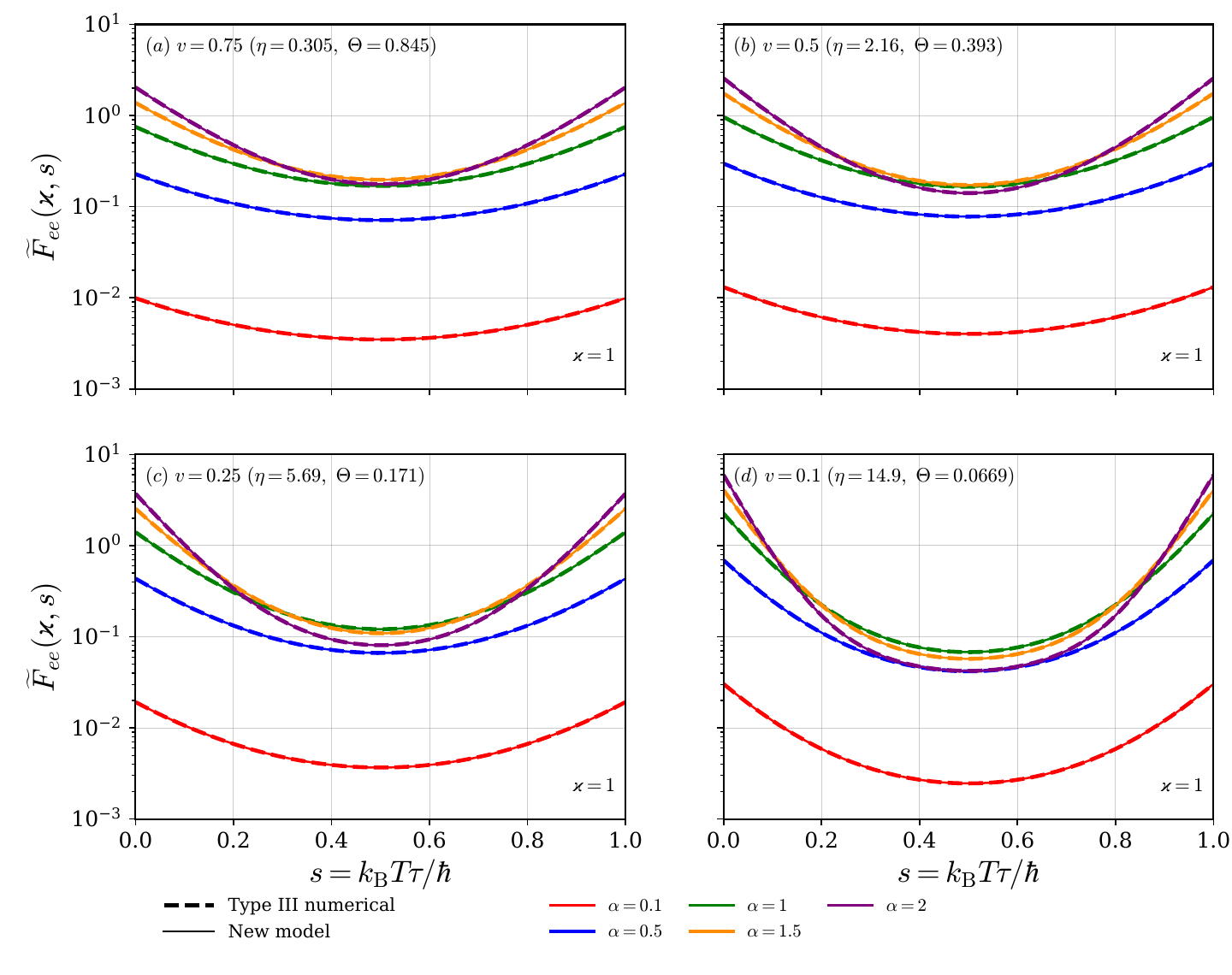}
	\caption{
		As in \fig{fig:itcf_comparison_kap=0.1}, but for $\varkappa=1$.
	}
	\label{fig:itcf_comparison_kap=1}
\end{figure}

The same spectral information can be represented by the imaginary-time correlation function (ITCF),
\begin{align}
	\label{eq:F_def}
	F_{ee}(k,\tau)
	= &\,
	\int_{-\infty}^{\infty}
	\dif\omega\,
	e^{-\omega\tau}
	S_{ee}(k,\omega)
	\,.
\end{align}
The ITCF arises naturally in finite-temperature path-integral calculations and has become increasingly important for model-free interpretation of XRTS measurements, including temperature determination, absolute-intensity normalisation and extraction of equilibrium properties \cite{Dornheim_NatCommun_2022, Dornheim_PhysPlasmas_2023b, Dornheim_PhilTransRoyalSocA_2023, Dornheim_MatterRadiatExtremes_2023, Dornheim_SciRep_2024, Dornheim_PhysPlasmas_2025, Bohme_PhysPlasmas_2026, Gawne_RevModPlasmaPhys_2026}.

Using detailed balance and introducing the dimensionless imaginary time
\begin{align}
	s=\frac{k_{\rm B}T\tau}{\hbar},
	\qquad 0\leq s\leq1,
\end{align}
a dimensionless representation equivalent to \eq{eq:dsf_def} is obtained:
\begin{align}
	\label{eq:F_def_2}
	\tilde{F}_{ee}(\varkappa,s;\alpha)
	\equiv &\,
	\frac{\pi n_e e^{2}}{\epsilon_{0}k^{2}}
    \frac{1}{4k_{\rm B}T}
	F_{ee}(k,\tau)
	\nnn
	= &\,
	\int_0^\infty
	\dif w\,
	\frac{L(\varkappa,w;\alpha)}
	{1-e^{-4w}}
	\nnn &\,
	\times
	\sqr{
		e^{-4sw}
		+
		e^{-4(1-s)w}
	}
	\,.
\end{align}
This expression satisfies the exact symmetry $\tilde{F}_{ee}(\varkappa,s;\alpha) = \tilde{F}_{ee}(\varkappa,1-s;\alpha)$, so that the ITCF is symmetric about $s=1/2$ \cite{Dornheim_NatCommun_2022,Dornheim_PhysPlasmas_2023b}.

The ITCF provides a complementary and, in some respects, more physically robust assessment of the approximation. The frequency integration in \eq{eq:F_def_2} suppresses sensitivity to small local shifts in narrow spectral features while retaining their integrated spectral weight. Consequently, accurate plasmon positions and integrated spectral weights are considerably more important for the ITCF than exact pointwise agreement at the maximum of an extremely narrow resonance.

As demonstrated in \fig{fig:itcf_comparison_kap=0.1} and \fig{fig:itcf_comparison_kap=1}, the new model reproduces the numerical ITCF with remarkable accuracy throughout the investigated parameter space. The maximum pointwise relative error remains below $0.166\%$, with 38/40 cases remaining below $0.1\%$. For $\varkappa=1$, the largest error is only $0.0525\%$. Even in the most strongly degenerate and weakly damped cases, the integrated response remains closely matched to the numerical benchmark.

\begin{figure}[!t]
	\centering
	\includegraphics[width=\columnwidth]
	{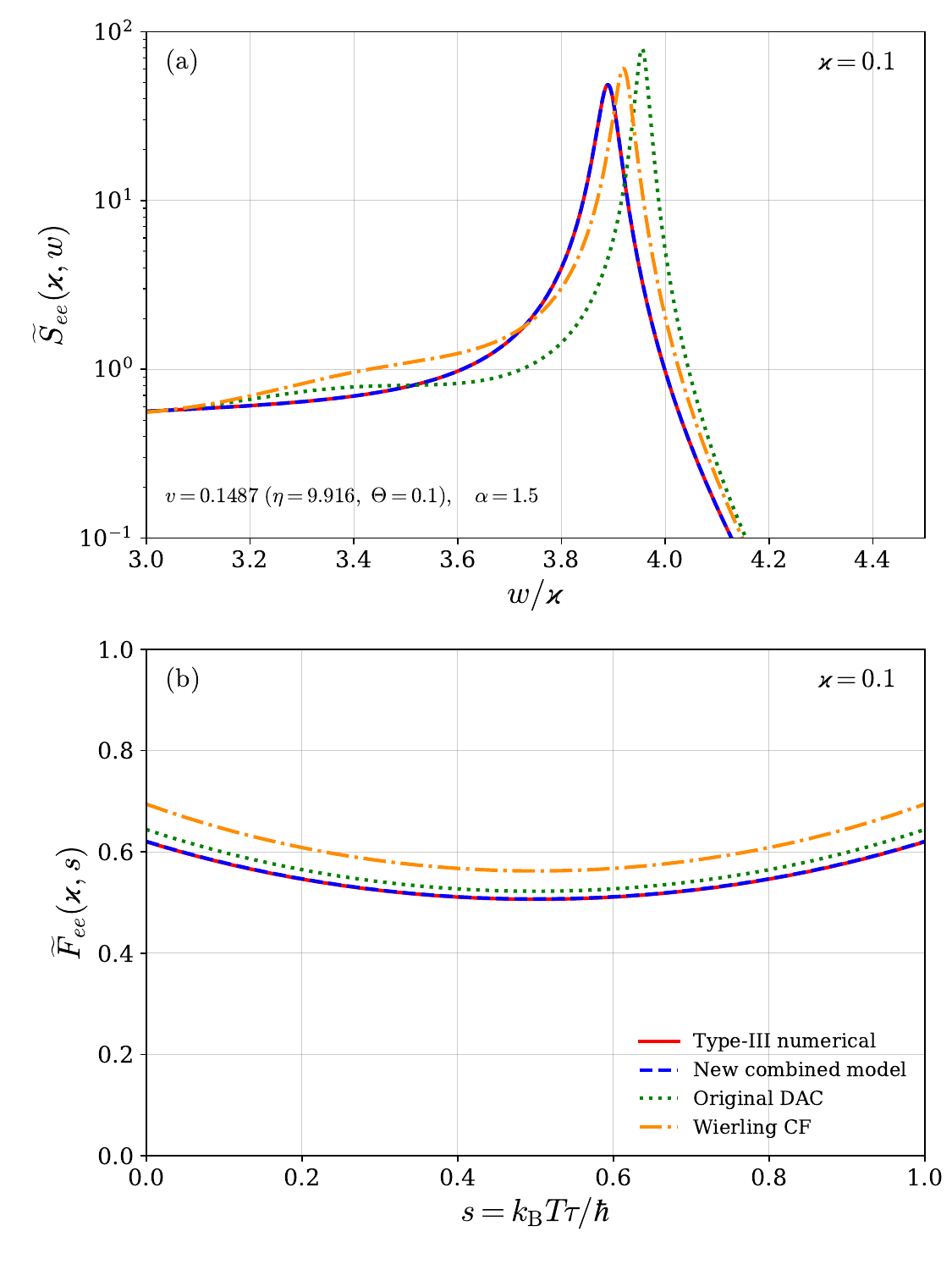}
	\caption{
		Comparison of the reduced DSF
		(a) and corresponding reduced ITCF
		(b) at $\Theta=0.1$ ($\eta=9.9164$, $v=0.148713$), $\varkappa=0.1$ and $\alpha=1.5$. The DSF is shown in the vicinity of the plasmon resonance, while the ITCF is obtained by integrating the complete corresponding spectrum. The new model closely reproduces both the resonance and its integrated imaginary-time response, whereas the older approximations produce appreciable discrepancies.
	}
	\label{fig:DSF_ITCF_model_comparison_Theta=0p1}
\end{figure}

The improvement obtained from the composite model is especially clear when compared directly with existing approximations. \fig{fig:DSF_ITCF_model_comparison_Theta=0p1} presents the DSF and corresponding ITCF at challenging conditions of partial degeneracy in the collective scattering regime: $\Theta=0.1$, $\eta=9.9164$, $v=0.148713$, $\varkappa=0.1$ and $\alpha=1.5$. The new model reproduces the numerical plasmon resonance extremely closely, whereas both the original DAC $4/5$ rational approximation and Wierling's continued fraction exhibit appreciable differences in resonance position and amplitude.

For this representative state, the numerical DSF peak amplitude is approximately $48.539$, compared with $48.531$ for the new model, $79.346$ for DAC and $60.537$ for Wierling. The corresponding relative differences in peak height are $-0.0169\%$, $+63.5\%$ and $+24.7\%$, respectively. These spectral differences propagate directly into the ITCF. At $s=1/2$, the numerical benchmark gives $\tilde{F}_{ee}=0.506477$, compared with $0.506980$ for the new model, $0.522304$ for DAC and $0.562404$ for Wierling. The respective relative differences are $+0.0993\%$, $+3.13\%$ and $+11.04\%$. Thus, the new model retains the integrated response to approximately one part in a thousand, whereas the older approximations introduce errors of order $10\%$ under identical physical conditions.

These comparisons highlight an important distinction between static fitting accuracy and the accuracy of derived dynamical observables. Although the new model was constructed primarily to reproduce $\HcalR$ and its scaled derivative $\Ical$, the resulting approximation also accurately describes the collective features of the dynamic dielectric response. The present results therefore provide strong independent support for the combined rational--polynomial formulation.

Further improvements could be obtained by incorporating the loss function or DSF directly into a multi-objective optimisation. Such a procedure could preferentially suppress residual errors near weakly damped resonances, where small errors in the real dielectric response have the greatest spectral consequences. The present results, however, indicate that this additional optimisation is not essential for obtaining accurate ITCF predictions.

The computational efficiency of the new representation is particularly useful for XRTS inference, where the dielectric response may be evaluated repeatedly during nonlinear regression, Bayesian sampling, uncertainty quantification and related inverse-problem workflows \cite{Bellenbaum_PhysPlasmas_2026, Poole_PhysPlasmas_2022, Poole_PhysRevRes_2024, Stanek_PhysPlasmas_2024, Allison_PhysPlasmas_2026}. The $6/7$ rational branch provides an especially inexpensive representation at lower degeneracy, while the composite polynomial branch extends the same accuracy advantages into the strongly degenerate regime without requiring repeated quadrature of $\HcalR$.

Finally, the present formulation offers an efficient route to constructing a dedicated approximation to the ITCF itself. Dense training data in $(\varkappa,\eta,\alpha,s)$ can be generated from the combined model, after which $\tilde{F}_{ee}(\varkappa,s;\alpha)$ may be parametrised directly. Its smoothness, exact symmetry about $s=1/2$, and relative insensitivity to narrow spectral resonances make such a reduced representation particularly attractive. A direct ITCF surrogate would eliminate both the dynamic dielectric reconstruction and frequency integration at runtime, offering a potentially useful acceleration for large-scale XRTS inference and related applications.

%% file: sections/main_body/07_conclusions.tex
\section{Conclusions}
\label{sec:conclusions}

A new global approximation has been developed for the real boundary of a regularised causal complex auxiliary function governing the static and dynamic dielectric response of the ideal electron gas within the random phase approximation. The formulation combines a $6/7$ rational representation for weakly and moderately degenerate conditions with a composite polynomial representation in the strongly degenerate regime, joined at $v=0.2$, where $v = \Fcal_{-1/2}(\eta)/\Fcal_{1/2}(\eta)$ is a compactified degeneracy coordinate. The rational model incorporates exact asymptotic constraints, doubly compactified coordinates and branch-locked optimisation of its coefficient functions, with simple polynomial parametrisation providing smooth degeneracy dependence and Bernstein-basis constraints ensuring a strictly positive denominator. In the high-degeneracy regime, the composite representation explicitly isolates the Kohn contribution and employs smooth polynomial and Chebyshev approximations to its regular remainder, retaining analytical integrability and continuity across the constituent regions.

Comparison with high-precision numerical integration demonstrates a substantial improvement over existing rational and continued-fraction approximations across all practical values of the function arguments. The new model reproduces both $\HcalR$ and its scaled derivative $\Ical = -\chi \partial \HcalR/\partial\chi$ with high accuracy, including the challenging strongly degenerate regime around the Kohn feature. Over the full benchmark grid, the maximum relative error in $\HcalR$ is approximately $0.109\%$, with the largest discrepancies concentrated near the transition between the two representations. Efficient Horner and Clenshaw evaluation further reduces the numerical cost without compromising accuracy. The median execution times are approximately $10$ and $63\,\mu$s per degeneracy state for the rational and composite branches, respectively, including state preparation and evaluation over $1001$ points in the primary argument. The resulting model therefore remains computationally inexpensive compared with repeated numerical quadrature.

The accuracy of the combined representation is also retained when propagated through the dynamic dielectric response. The resulting electron--electron dynamic structure factor agrees closely with the numerical benchmark, reproducing the positions, widths and amplitudes of collective plasmon resonances across both degeneracy regimes. The vast majority of all cases examined lead to peak-height errors below $1\%$, with a maximum discrepancy of approximately $1.04\%$. Although narrow, weakly damped resonances remain particularly sensitive to small errors in the real dielectric response, the composite representation substantially alleviates the spectral inaccuracies of existing approximation schemes. The corresponding imaginary-time correlation function is reproduced with even greater fidelity: its maximum pointwise relative error remains below $0.166\%$ across all cases. Direct comparison with the original DAC rational approximation and Wierling's continued fractions further demonstrates the superiority of the new model in preserving the integrated spectral response under strongly degenerate conditions.

These results establish the combined representation as an accurate and efficient foundation for finite-wavelength screening calculations, dynamic dielectric modelling and computationally demanding X-ray Thomson scattering analysis. Its combination of high accuracy, regular numerical behaviour and inexpensive evaluation is particularly advantageous for inverse problems requiring repeated calculations over broad thermodynamic parameter ranges. Further improvements could be obtained by incorporating the loss function directly into a multi-objective optimisation, particularly to reduce residual errors near weakly damped resonances. The present formulation also provides a practical basis for constructing direct surrogate models of the imaginary-time correlation function, potentially eliminating the additional frequency integration during large-scale inference calculations.

\section*{Acknowledgements}

The author would like to thank Dr Iyyappa Rajan Panneerselvam and Dr James Allison for fruitful conversations and the Office of the CTO at First Light Fusion for internal review and supporting publication of this work.

\section*{Data availability}

All elements of the model required for reproduction of this work are documented herein.

%% file: sections/appendices/01_DAC_model_equations.tex
\section{Reformulation of original DAC model equations}
\label{app:DAC_model_equtions}

The full list of coefficients required to evaluate the rational functions used in the various approximations discussed in this work are listed in the following tables. The original DAC model uses the following parametrised functions (re-written here to align with the presentation of the values shown in \tab{tab:original_DAC_coefficients})
\begin{widetext}
	\begin{align}
		\label{eq:DAC_I_m0p5}
		I_{-1/2}(\Theta)
		= &\,
		\frac{2}{\Theta^{1/2}}
		\frac{1 + c_{0}\Theta^{2} + c_{1}\Theta^{4} + c_{2}\Theta^{6}}{1 + c_{3}\Theta^{2} + c_{4}\Theta^{4} + c_{5}\Theta^{11/2} + c_{6}\Theta^{7}}
		\nnn &
		\Rightarrow
		\sqrt{\pi}\Fcal_{-1/2}(\eta)
		\,,
		\\
		I_{3/2}(\Theta)
		= &\,
		\frac{2}{5\Theta^{5/2}}
		\frac{1 + c_{0}\Theta^{2} + c_{1}\Theta^{4} + c_{2}\Theta^{8}}{1 + c_{3}\Theta^{2} + c_{4}\Theta^{4} + c_{5}\Theta^{11/2} + c_{6}\Theta^{7}}
		\nnn &
		\Rightarrow
		\frac{3\sqrt{\pi}}{4}\Fcal_{3/2}(\eta)
		\,,
		\\
		I_{5/2}(\Theta)
		= &\,
		\frac{2}{7\Theta^{7/2}}
		\frac{1 + c_{0}\Theta^{2} + c_{1}\Theta^{4} + c_{2}\Theta^{15/2}}{1 + c_{3}\Theta^{2} + c_{4}\Theta^{4} + c_{5}\Theta^{11/2}}
		\nnn &
		\Rightarrow
		\frac{15\sqrt{\pi}}{8}\Fcal_{5/2}(\eta)
		\,,
		\\
		\label{eq:DAC_J}
		J(\Theta)
		= &\,
		\frac{1 + c_{0}\Theta^{2} + c_{1}\Theta^{4} + c_{2}\Theta^{7}}{1 + c_{3}\Theta^{2} + c_{4}\Theta^{4} + c_{5}\Theta^{6} + c_{6}\Theta^{15/2} + c_{7}\Theta^{9}}
		\nnn &
		\Rightarrow
		\sqrt{\frac{\pi}{\Theta}}\,
		\FD{-3/2}(\eta)
		\,,
		\\
		\label{eq:DAC_K}
		K(\Theta)
		= &\,
		\frac{1 + c_{0}\Theta^{2} + c_{1}\Theta^{4} + c_{2}\Theta^{7}}{1 + c_{3}\Theta^{2} + c_{4}\Theta^{4} + c_{5}\Theta^{7} + c_{6}\Theta^{17/2} +  c_{7}\Theta^{10}}
		\nnn &
		\Rightarrow
		-\frac{2}{\Theta}
		\sqrt{\frac{\pi}{\Theta}}\,
		\FD{-5/2}(\eta)
		\,,
		\\
		\label{eq:DAC_a_2}
		a_{2}(\Theta)
		= &\,
		\frac{c_{0} + \Theta}{c_{1} + c_{2}\Theta^{c_{3}} + c_{4}\Theta^{2}}
		\,,
		\\
		\label{eq:DAC_a_4}
		a_{4}(\Theta)
		= &\,
		\frac{1 + c_{0}\Theta + c_{1}\Theta^{2}}{c_{2} + c_{3}\Theta + c_{4}\Theta^{2} + c_{5}\Theta^{3} + c_{6}\Theta^{4}}
		\,,
		\\
		\label{eq:DAC_a_6}
		a_{6}(\Theta)
		= &\,
		\frac{c_{0} + \Theta}{c_{1} + c_{2}\Theta + c_{3}\Theta^{2} + c_{4}\Theta^{3} + c_{5}\Theta^{4}}
		\,,
		\\
		\label{eq:DAC_a_8}
		a_{8}(\Theta)
		= &\,
		\frac{c_{0} + c_{1}\Theta + c_{2}\Theta^{2}}{1 + c_{3}\Theta + c_{4}\Theta^{2} + c_{5}\Theta^{3} + c_{6}\Theta^{4} + c_{7}\Theta^{5} + c_{8}\Theta^{6}}
		\,,
		\\
		\label{eq:DAC_eta}
		\eta(\Theta)
		= &\,
		\exp\left[
			-c_{0}\pi^{2}\Theta^{2}\rnd{1 + c_{1}\pi^{2}\Theta^{2}}
		\right]
		-
		c_{2}
		\ln\rnd{\frac{3\sqrt{\pi}}{4}\Theta^{3/2}}
		\exp\sqr{
			-\rnd{
				1 + \frac{c_{3}}{\pi^{2} \Theta^{2}} + \frac{c_{4}}{\pi^{4}\Theta^{4}}
			}
		}
		\,,
	\end{align}
\end{widetext}
The last equation comes from Dharma-wardana and Taylor's fit \cite{Dharma-wardana_JPhysC_1981}, which is stated as the preferred model by DAC.

%% file: sections/appendices/02_end_point_construction.tex
\section{Endpoint construction of the $6/7$ rational approximation}
\label{app:H67_endpoint_constraints}

The denominator coefficients of the $6/7$ rational approximation are determined algebraically by matching the exact short- and long-wavelength asymptotic expansions of $\HcalR$ through cubic order in the compactified coordinate $u$. This eliminates all eight Bernstein coefficients as independent optimisation parameters, leaving only six numerator-controlling functions to be fitted.

Using the compactified coordinate $u=t/(1+t)$ with $t = 2v\chi^2$, the approximation takes the form
\begin{align}
	\HcalR(\chi,\eta)
	\simeq
	(1-u)\frac{P_6(u;v)}{Q_7(u;v)},
\end{align}
where
\begin{align}
	P_6(u;v)
	= &\,
	1+\sum_{j=1}^{6}\alpha_j(v)u^j,
	\\
	Q_7(u;v)
	= &\,
	\sum_{j=0}^{7}
	\zeta_j(v)\binom{7}{j}
	u^j(1-u)^{7-j}.
\end{align}
The denominator is expressed in the Bernstein basis, which facilitates verification of polynomial positivity \cite{Farouki_ComputAidedGeom_2012}.

The exact long-wavelength expansion is
\begin{align}
	\HcalR(t\rightarrow0,\eta)
	=
	1+c_1t+c_2t^2+c_3t^3
	+\Ocal(t^4),
\end{align}
with
\begin{align}
	c_1
	= &\,
	-\frac{\Gcal_{-3/2}}{3v^2}
	\,,
	\nnn
	c_2
	= &\,
	\frac{\Gcal_{-5/2}}{15v^3}
	\,,
	\nnn
	c_3
	= &\,
	-\frac{\Gcal_{-7/2}}{105v^4}
	\,.
\end{align}
Here $\Gcal_j(\eta)=\mathcal{F}_j(\eta)/ \mathcal{F}_{1/2}(\eta)$, as in the main text. In the opposite limit,
\begin{align}
	\HcalR(t\rightarrow\infty,\eta)
	=
	\frac{1}{t}
	+\frac{r_2}{t^2}
	+\frac{r_3}{t^3}
	+\frac{r_4}{t^4}
	+\Ocal(t^{-5})
	\,,
\end{align}
where
\begin{align}
	r_2
	= &\,
	v\, \Gcal_{3/2}
	\,,
	\nnn
	r_3
	= &\,
	3v^2\, \Gcal_{5/2}
	\,,
	\nnn
	r_4
	= &\,15v^3\, \Gcal_{7/2}
	\,.
\end{align}

Defining the residual rational function $R_{6/7}=P_6/Q_7$, the transformation $t=u/(1-u)$ gives the endpoint expansions
\begin{align}
	R_{6/7}(u\rightarrow0)
	= &\,
	1+\lambda_1u+\lambda_2u^2
	+\lambda_3u^3+\Ocal(u^4)
	\,,
	\\
	R_{6/7}(u\rightarrow1)
	= &\,
	1+\rho_1z+\rho_2z^2
	+\rho_3z^3+\Ocal(z^4)
	\,,
\end{align}
where $z=1-u$, and
\begin{align}
	\lambda_1 
	= 
	&\,1+c_1
	\,,
	&
	\rho_1 
	= 
	&\,1+r_2
	\,,
	\nnn
	\lambda_2 
	= 
	&\,1+2c_1+c_2
	\,,
	&
	\rho_2 
	= 
	&\,1+2r_2+r_3
	\,,
	\nnn
	\lambda_3 
	= 
	&\,1+3c_1+3c_2+c_3
	\,,
	&
	\rho_3 
	= 
	&\,1+3r_2+3r_3+r_4
	\,.
\end{align}

The denominator is obtained directly from
$Q_7=P_6/R_{6/7}$. At $u=0$, writing
\begin{align}
	Q_7(u)
	=
	A_0+A_1u+A_2u^2+A_3u^3
	+\Ocal(u^4)
	\,,
\end{align}
series division gives
\begin{align}
	A_0 
	= 
	&\,1
	\,,
	\nnn
	A_1 
	= 
	&\,\alpha_1-\lambda_1
	\,,
	\nnn
	A_2 
	= &\,
	\alpha_2 - \lambda_1 \alpha_1 + \lambda_1^2 - \lambda_2
	\,,
	\nnn
	A_3 
	= 
	&\,\alpha_3 - \lambda_1 \alpha_2 + (\lambda_1^2 - \lambda_2) \alpha_1
	\nnn &
	-
	\lambda_1^3 + 2\lambda_1 \lambda_2 - \lambda_3
	\,.
\end{align}
Comparison with the Bernstein expansion determines the first four denominator coefficients:
\begin{align}
	\zeta_0 
	= &\,
	1\,,
	\nnn
	\zeta_1 
	= 
	&\,
	\zeta_0+\frac{A_1}{7}
	\,,
	\nnn
	\zeta_2 
	= &\,
	2\zeta_1-\zeta_0+\frac{A_2}{21}
	\,,
	\nnn
	\zeta_3 
	= 
	&\,3\zeta_2-3\zeta_1+\zeta_0 + \frac{A_3}{35}
	\,.
\end{align}

At $u=1$, define the numerator endpoint value and derivatives
\begin{align}
	P_* = &\,P_6(1)
	=1+\sum_{j=1}^{6}\alpha_j,
	\\
	P_*^{(n)} = &\,P_6^{(n)}(1)
	=\sum_{j=n}^{6}
	\frac{j!}{(j-n)!}\alpha_j
	\,,
\end{align}
for $n=1,2,3$. Expanding the denominator in powers of $z=1-u$,
\begin{align}
	Q_7(1-z)
	=
	B_0+B_1z+B_2z^2+B_3z^3
	+\Ocal(z^4),
\end{align}
gives
\begin{align}
	B_0 = &\,P_*,
	\nnn
	B_1 = &\,-P_*^{(1)}-\rho_1P_*,
	\nnn
	B_2 = &\,\frac{P_*^{(2)}}{2}
	+\rho_1P_*^{(1)}
	+(\rho_1^2-\rho_2)P_*,
	\nnn
	B_3 = &\,-\frac{P_*^{(3)}}{6}
	-\frac{\rho_1P_*^{(2)}}{2}
	\nnn
	&-(\rho_1^2-\rho_2)P_*^{(1)}
	\nnn
	&+(-\rho_1^3+2\rho_1\rho_2-\rho_3)P_*.
\end{align}
The remaining four Bernstein coefficients follow as
\begin{align}
	\zeta_7 = &\,B_0=P_*,
	\nnn
	\zeta_6 = &\,\zeta_7+\frac{B_1}{7},
	\nnn
	\zeta_5 = &\,2\zeta_6-\zeta_7+\frac{B_2}{21},
	\nnn
	\zeta_4 = &\,3\zeta_5-3\zeta_6+\zeta_7
	+\frac{B_3}{35}.
\end{align}

Consequently, all eight denominator coefficients
$\zeta_0,\ldots,\zeta_7$ are determined exactly by the numerator coefficients and the known asymptotic moments. No independent denominator functions are required.

In the final parametrisation, positivity of the denominator at $u=1$ is incorporated explicitly by defining
\begin{align}
	\xi(v)
	\equiv\ln P_6(1;v)
	=\ln\zeta_7(v).
\end{align}
The six independent state functions are therefore $\{\alpha_1,\ldots,\alpha_5,\xi\}$, with
\begin{align}
	\alpha_6(v)
	=
	e^{\xi(v)}
	-1-\sum_{j=1}^{5}\alpha_j(v).
\end{align}
These functions are represented using Chebyshev polynomials. The remaining coefficients are reconstructed algebraically at each state, preserving the exact endpoint constraints and their smooth dependence on $v$.

The condition $\zeta_7=e^\xi>0$ guarantees positivity at $u=1$, but does not by itself guarantee $Q_7(u)>0$ throughout the interval. Interior positivity is therefore imposed and verified separately using Bernstein-basis subdivision constraints during optimisation and validation.

%% file: sections/appendices/03_branch_locked_fitting.tex
\section{Predictor--corrector branch locking for statewise coefficient optimisation}
\label{app:branch_locked_fitting}

The statewise optimisation of the rational coefficients can possess nearly degenerate directions, allowing distinct coefficient vectors to reproduce the same response with comparable accuracy. Independent optimisation at successive values of $v$ may therefore switch between equivalent minima, introducing non-physical oscillations that complicate subsequent polynomial parametrisation. To maintain a smoothly connected coefficient sequence, we employ a branch-locked predictor--corrector continuation scheme with adaptive stepping, following the general methodology of Rheinboldt \cite{Rheinboldt1980} and Allgower and Georg \cite{AllgowerGeorg1990}.

At each proposed state $v_{j+1}=v_j+\Delta v$, the coefficient vector is predicted by extrapolation along the previously accepted branch,
\begin{align}
	\cvec_{j+1}^{\,\mathrm{pred}}
	=\cvec_j+
	\left(\frac{d\cvec}{dv}\right)_j\Delta v,
\end{align}
where the local gradient is estimated from preceding states. The corrector optimisation is then initialised near this prediction and accepted only if
\begin{align}
	\max_i
	\frac{|c_i-c_i^{\mathrm{pred}}|}
	{m\,s_i+C_i e_{i,\mathrm{pred}}}
	\leq 1
	\,,
	\\
	e_{i,\mathrm{pred}}
	\simeq
	\frac{1}{2}
	\left|\frac{d^2c_i}{dv^2}\right|
	(\Delta v)^2.
\end{align}
Here $s_i$ denotes a coefficient-specific numerical scale, estimated from stable branch segments and repeated fixed-state optimisations; $m$ controls the branch-lock strength, while $C_i$ weights the local curvature allowance. This criterion accommodates smooth physical variation without permitting arbitrary jumps between neighbouring minima.

If the acceptance condition fails, the proposed step is reduced, typically by $\Delta v\rightarrow\Delta v/2$, and the prediction and optimisation are repeated without relaxing the branch-lock tolerance. Absolute, rather than coefficient-relative, scales are employed to avoid artificial restrictions near zero crossings. The resulting continuation selects a smooth representative of the nearly degenerate solution family, suppressing optimisation-induced oscillations and facilitating low-order parametrisation of the coefficient functions.

%% file: sections/appendices/04_hermite_interpolant.tex
\section{Quintic Hermite interpolant construction}
\label{app:hermite_interpolant}

The regular polynomial contribution within the Kohn window $u_{<}\leq u\leq u_{>}$ is constructed to reproduce the value, first derivative and curvature of the logarithmically divergent contribution at both boundaries. With the local coordinate $\tilde{u}_{2}$ \eqref{eq:tilde_u_2} and region width \eqref{eq:Delta_u_K}, the divergent contribution is
\begin{align}
	\Ical_{\Kcal}^{\rm div}(\chi_0)
	=\frac{1+\chi_0^2}{4\chi_0}
	\ln|1-\chi_0|
	\,,
\end{align}
with $\chi_0(u) = \sqrt{u/2v\eta(1-u)}$. For the polynomial representation of the regular Kohn contribution, define
\begin{align}
	Y(u)
	=-\frac{\Ical_{\Kcal}^{\rm div}(\chi_0(u))}
	{u(1-u)}.
	\label{eq:kohn_hermite_target}
\end{align}
The endpoint values and derivatives are evaluated analytically, with differentiation taken with respect to $u$. In particular, writing $f(u)=u(1-u)$ and $D(u)=\Ical_{\Kcal}^{\rm div}(\chi_0(u))$, one has
\begin{align}
	Y' &=
	\frac{-D'-f'Y}{f},
	\\
	Y'' &=
	\frac{-D''+2Y-2f'Y'}{f},
\end{align}
where $f'=1-2u$. The required derivatives of $D$ follow from the chain rule, using
\begin{align}
	\frac{\dif\chi_0}{\dif u}
	&=\frac{\chi_0}{2u(1-u)},
	\\
	\frac{\dif^2\chi_0}{\dif u^2}
	&=\frac{\chi_0(u-1/4)}{u^2(1-u)^2}.
\end{align}

Let $Y_{\pm}$, $Y'_{\pm}$ and $Y''_{\pm}$ denote the corresponding quantities evaluated at $u_{<}$ and $u_{>}$. The dimensionless endpoint slopes and curvatures are
\begin{align}
	m_{\pm} &=\Delta u\,Y'_{\pm},
	&
	q_{\pm} &=(\Delta u)^2Y''_{\pm}.
\end{align}
The unique quintic Hermite interpolant is written
\begin{align}
	h_5(\tilde{u}_2)
	=\sum_{j=0}^{5}c_j\tilde{u}_2^j,
\end{align}
where, defining $\Delta Y=Y_{>}-Y_{<}$,
\begin{align}
	c_0={}&Y_{<},
	\\
	c_1={}&m_{<},
	\\
	c_2={}&\frac{q_{<}}{2},
	\\
	c_3={}&10\Delta Y-6m_{<}-4m_{>}
	-\frac{3q_{<}}{2}+\frac{q_{>}}{2},
	\\
	c_4={}&-15\Delta Y+8m_{<}+7m_{>}
	+\frac{3q_{<}}{2}-q_{>},
	\\
	c_5={}&6\Delta Y-3m_{<}-3m_{>}
	-\frac{q_{<}}{2}+\frac{q_{>}}{2}.
\end{align}
By construction, $h_5$ reproduces $Y$, $Y'$ and $Y''$ at both boundaries, ensuring exact $C^2$ matching.

The complete regular contribution within the Kohn window is then represented as
\begin{align}
	\Ical_{\Kcal}^{\rm reg}(u;v,\eta)
	={}&u(1-u)\Big[
	h_5(\tilde{u}_2)
	\nnn &+
	\tilde{u}_2^3(1-\tilde{u}_2)^3
	P_{4,4}(\tilde{u}_2,v)
	\Big],
\end{align}
where $P_{4,4}$ denotes the fitted bivariate Chebyshev polynomial. The cubic factors guarantee that the fitted correction and its first two derivatives vanish at both boundaries, preserving the Hermite matching conditions.

The resulting expression is polynomial in $\tilde{u}_2$ and can be integrated analytically to obtain its contribution to $\HcalR$. Thus, the boundary-matching coefficients are derived directly from the divergent Kohn contribution at each thermodynamic state, without introducing additional optimisation parameters or numerical quadrature.

%% file: sections/appendices/05_fitting_coefficients.tex
\begin{widetext}	

\section{Numerical coefficients of the original DAC, hybrid and new models}
\label{app:coefficients}

\begin{table*}[!h]
	\centering
	\small
	\begin{tabular}{lrrrrrrrrr}
		\hline
		Func.
		& $m = 0$ 
		& $m = 1$ 
		& $m = 2$ 
		& $m = 3$ 
		& $m = 4$
		& $m = 5$ 
		& $m = 6$ 
		& $m = 7$ 
		& $m = 8$ 
		\\
		\hline
		$I_{-1/2}$
		& $41.775$ 
		& $27.390$ 
		& $4287.2$ 
		& $c_{0} + \frac{1}{12}\pi^2$ 
		& $50.605$ 
		& $\frac{1}{\sqrt{2\pi}}c_{2}$ 
		& $\frac{3}{2}c_{2}$ 
		& --
		& --
		\\
		$I_{3/2}$
		& $5.3588$  
		& $-2.5433$
		& $432.89$
		& $c_{0} - \frac{5}{12}\pi^{2}$
		& $1.88$ 
		& $-\frac{2}{15\sqrt{2\pi}}c_{2}$ 
		& $\frac{2}{5}c_{2}$
		& --
		& --
		\\
		$I_{5/2}$
		& $-8.6164$ 
		& $-357.41$ 
		& $5711.1$
		& $c_{0} - \frac{7}{6}\pi^{2}$ 
		& $-\frac{1}{35\sqrt{\pi/2}}c_{2}$ 
		& $\frac{4}{35}c_{2}$ 
		& --
		& --
		& --
		\\
		$J$
		& $3248.8$ 
		& $-691.47$
		& $-3202700$ 
		& $c_{0} - \frac{1}{6}\pi^2$
		& $-4535.6$
		& $-462400$ 
		& $\frac{3}{4\sqrt{\pi/2}}c_{2}$ 
		& $\frac{3}{4}c_{2}$
		& --
		\\
		$K$
		& $-4.8780$ 
		& $473.25$
		& $-2337.5$ 
		& $c_{0} - \frac{3}{4}\pi^2$
		& $348.31$
		& $1517.3$ 
		& $-\frac{7}{8\sqrt{\pi/2}}c_{2}$
		& $-\frac{3}{8}c_{2}$
		& --
		\\
		$a_2$
		& $-0.22800$ 
		& $0.42220$ 
		& $-0.64660$
		& $0.70572$ 
		& $5.8820$ 
		& -- 
		& -- 
		& -- 
		& -- 
		\\
		$a_4$
		& $-3.0375$ 
		& $64.646$ 
		& $19.608$ 
		& $-96.978$
		& $423.66$ 
		& $-331.01$ 
		& $20.833\,c_{1}$ 
		& -- 
		& -- 
		\\
		$a_6$
		& $-0.19$ 
		& $0.36538$ 
		& $-2.2575$
		& $22.942$ 
		& $-43.492$ 
		& $106.40$ 
		& --
		& --
		& --
		\\
		$a_8$
		& $0.91$
		& $-6.4453$ 
		& $12.2324$ 
		& $-7.1316$ 
		& $22.725$ 
		& $58.092$ 
		& $-436.02$
		& $-826.51$ 
		& $4912.9$ 
		\\
		$\eta$
		& $1/12$
		& $0.178$ 
		& $2.718$ 
		& $1.75$ 
		& $59.4$ 
		& --
		& --
		& --
		& --
		\\
		\hline
	\end{tabular}
	\caption{
		Numerical constants for the original DAC Pad\'{e} model. The coefficients $c_m$ are defined separately for each function and are re-written here to improve clarity for implementation. The FDI rows correspond to Eqs.~(A1)--(A4) of Ref.~\cite{Dandrea_PhysRevB_1986}; the remaining rows correspond to the functions fitted for the implementation of the rational approximation to $\HcalR$, Eqs.~(A19)--(A24) of Ref.~\cite{Dandrea_PhysRevB_1986}. The table of constants given here is to be used in conjunction with the re-written DAC model expressions \eqs{eq:DAC_I_m0p5}{eq:DAC_a_8} of this work.
	}
	\label{tab:original_DAC_coefficients}
\end{table*}

\begin{table*}[t]
	\centering
	\small
	\begin{tabular}{lrrrrr}
		\hline
		Func. 
		& $m = 0$ 
		& $m = 1$ 
		& $m = 2$ 
		& $m = 3$ 
		& $m = 4$ 
		\\
		\hline
		$P_2$ 
		& $ 0.4247654212732541$ 
		& $-0.2489665876936149$
		& $-1.797149480935104$  
		& $ 2.564301940597440$
		& $-0.9458489200507652$ 
		\\
		$P_4$ 
		& $ 0.3086267320441940$ 
		& $-0.9682399344018132$
		& $ 1.275848843903844$  
		& $-0.8799560839672742$
		& $ 0.2662839557612876$ 
		\\
		$P_6$ 
		& $ 0.06345806239987775$ 
		& $ 0.07467351421561658$
		& $-0.7053650085611638$ 
		& $ 0.9281681278053068$
		& $-0.3615665244877381$ 
		\\
		$P_8$ 
		& $ 0.1176238337677901$ 
		& $-0.5248748036916447$
		& $ 1.104367180578949$  
		& $-1.087948048805761$
		& $ 0.3913865593330271$ 
		\\
		$Q_2$ 
		& --
		& $-1.579009419090817$ 
		& $-0.9483903885921836$
		& $ 2.789739772163082$ 
		& $-1.256543737560078$ 
		\\
		$Q_4$ 
		& --
		& $-3.873658103573866$ 
		& $ 6.161165281249496$
		& $-4.701355914905370$ 
		& $ 1.447010075000919$ 
		\\
		$Q_6$ 
		& --
		& $-3.999999991755386$ 
		& $ 6.119815107477399$
		& $-4.235001940311175$ 
		& $ 1.115918898855183$ 
		\\
		$Q_8$ 
		& --
		& $-3.999999991755386$ 
		& $ 6.126961813657768$
		& $-4.243774318558717$ 
		& $ 1.117200474943613$ 
		\\
		\hline
	\end{tabular}
	\caption{
		Coefficients for the hybrid/re-optimised 4/5 rational fit of the DAC model.
	}
	\label{tab:hybrid_DAC_coefficients}
\end{table*}

\begin{table*}[t]
    \centering
    \scriptsize
    \setlength{\tabcolsep}{3pt}
    \renewcommand{\arraystretch}{1.12}
    \begin{tabular}{lrrrrr}
        \hline
        Func. & $m=0$ & $m=1$ & $m=2$ & $m=3$ & $m=4$ \\
        \hline
        $\gamma$ & 
        1.7892437422818492 & -0.01870117820907646 & 0.0099703028025722729 & -0.0059463127815412627 & 0.0043562102745323683 \\
        $s_1$ & 
        9.9651238486873499 & -0.45871215518747821 & 0.24093740935726699 & -0.14451150149107081 & 0.10592272532519556 \\
        $s_2$ & 
        -9.9781953361881044 & 0.69839730017588952 & -0.36037301760488238 & 0.21688784511391551 & -0.15933788364265322 \\
        $s_3$ & 
        5.003920954136138 & -0.47147337631188313 & 0.23998453925680732 & -0.14526050400914264 & 0.10692704852870977 \\
        $s_4$ & 
        -1.0058481805566721 & 0.12033585338099048 & -0.060964585229894529 & 0.037253501288130972 & -0.027372598775318308 \\
        $\xi$ & 
        -11.443081709585746 & -3.6378626535583409 & 0.70429262772212486 & -0.089255187576294179 & -0.23983296410449823 \\
        \hline
    \end{tabular}
    \par\vspace{0.75em}
    \begin{tabular}{lrrrrr}
        \hline
        Func. & $m=5$ & $m=6$ & $m=7$ & $m=8$ & $m=9$\\
        \hline
        $s_1$ & 
        -0.0031179716028450146 & 0.0013015462630573097 & -0.00053546105664013539 & 0.00066568625576934035 &
        0.00043354608308601304 \\
        $s_2$ & 
        0.0067997528959695209 & -0.0026516560456547066 & 0.0010272575263906825 & -0.0016932678259387079 &
        -0.0012484170767819559 \\
        $s_3$ & 
        -0.0060608342176158483 & 0.0023104779039408016 & -0.00086045795254409321 & 0.0015429155385024247 &
        0.0011704184755963901 \\
        $s_4$ & 
        0.0020778025718486224 & -0.00071896623301499123 & 0.00018342879439335754 & -0.0004081764489950376 &
        -0.00037700111750923722 \\
        $\xi$ & 
        -0.011939860077315644 & -0.17319183200409366 & -0.0097050050438698809 & 0.090521172905011174 &
        0.041404358262493959 \\
        \hline
    \end{tabular}
    \caption{
        Chebyshev coefficients $A_{m}$, $B_{nm}$ and $C_{m}$ used in the $6/7$ rational approximation for the low-degeneracy regime $v_{0} \leq v < v_{\rm class}$.
    }
    \label{tab:global_cheb55_coefficients}
\end{table*}

\begin{table*}[t]
\centering
\small
\setlength{\tabcolsep}{4pt}
\renewcommand{\arraystretch}{1.12}
\begin{tabular}{rrrrr}
\hline
Func. & $m=0$ & $m=1$ & $m=2$ & $m=3$ \\
\hline
$u_{<}$ & $-0.28349958886903925$ & $-0.46227446243230236$ & $-0.47694961017796156$ & $-1.0138537001113404$ \\
$u_{>}$ & $0.28356892616565604$ & $-0.4592064629977331$ & $0.5419498082147971$ & $-0.71858918560783436$ \\
\hline
\end{tabular}
\caption{
	Quartic coefficients $D_{m}^{\gtrless}$ for the Kohn-region endpoints $u_{\gtrless}=\frac{3}{4}+\sum_{m=0}^3 D_{m}^{\gtrless}v^{m+1}$. The half-width of the smooth-background transition is derived from these endpoints and requires no additional fitted coefficients.
}
\label{tab:new_kohn_boundary_coefficients}
\end{table*}

\begin{table*}[t]
	\centering
	\scriptsize
	\setlength{\tabcolsep}{4pt}
	\renewcommand{\arraystretch}{1.12}
	\begin{tabular}{rrrrr}
		\hline
		$n$ & $m = 0$ & $m = 1$ & $m = 2$ & $m = 3$ \\
		\hline
		$0$  & $0.16456445123483784$     & $0.0038775923041734859$   & $-0.0019238949476440682$    & $-0.0008949454945255381$ \\
		$1$  & $0.023940289444507306$    & $0.020810594901387106$    & $-0.0014450353831802763$    & $-0.0011627504497523388$ \\
		$2$  & $0.088621973308633181$    & $0.0065646236240601424$   & $-0.0035962165693402544$    & $-0.0015482457367465543$ \\
		$3$  & $-0.028569353892370378$   & $-0.031333391606635543$   & $-0.0026780114550512659$    & $-0.0013748366295125406$ \\
		$4$  & $-0.042202593184897511$   & $0.0058113104260333398$   & $-0.0028883768236445235$    & $-0.0016489882437504399$ \\
		$5$  & $-0.0063565539452948546$  & $-0.0087845801564741706$  & $-0.0016414257024262423$    & $-0.00096563395150023974$ \\
		$6$  & $-0.018624046029085777$   & $0.014936632124634884$    & $-0.00028362638408739219$   & $-0.0016580769263598155$ \\
		$7$  & $0.013157371688229754$    & $0.0082437649096583909$   & $-0.0036428743496461995$    & $-0.0011260524105733928$ \\
		$8$  & $0.0061676959380281638$   & $0.0062068063230201692$   & $-0.0017254381834796778$    & $-0.0014615270805407327$ \\
		$9$  & $0.0094183110721248898$   & $0.003394784680129828$    & $-0.0050060741360877835$    & $-0.0010555422637330059$ \\
		$10$ & $0.0048122121330309141$   & $-0.0034741957769896929$  & $-0.0028651481699323023$    & $-0.00095348326393534773$ \\
		$11$ & $0.00259430392673226$     & $-0.0020699768250280327$  & $-0.0038948943440604089$    & $-0.00065297972981714859$ \\
		$12$ & $0.00010794812311685471$  & $-0.004717980988856894$   & $-0.0021016034833792369$    & $-0.00049881604756099975$ \\
		$13$ & $0.0011009349898223348$   & $-0.0019378763317346255$  & $-0.0023680023130437369$    & $-0.00032727196364711807$ \\
		$14$ & $-0.00019148279299702943$  & $-0.0032632732388875986$  & $-0.0011445771627850577$    & $-0.0002423043840049931$ \\
		$15$ & $0.00094657101063079854$  & $-0.00104463174035067$    & $-0.0014970801811335077$    & $-0.00014489728306453471$ \\
		$16$ & $0.00019526366673893009$  & $-0.0026281136858099913$  & $-0.00063672025563569503$   & $-6.7253551180238605\times10^{-5}$ \\
		$17$ & $0.0001718621490718967$   & $-0.00085650948035816794$ & $-0.00079163228972620517$   & $-3.0161267114685119\times10^{-5}$ \\
		$18$ & $-0.00011380143134915037$  & $-0.0017429981968302669$  & $-0.00023816076585018585$   & $2.6262644140507233\times10^{-5}$ \\
		$19$ & $-0.00016843883680075388$  & $-0.0004181873392543459$  & $-0.00019093024277190495$   & $8.3151305675064603\times10^{-6}$ \\
		$20$ & $-0.00017546117890987801$  & $-0.00055423952139844297$ & $-1.1331346642381572\times10^{-6}$ & $3.5766802404464104\times10^{-5}$ \\
		\hline
	\end{tabular}
	\caption{
		Chebyshev coefficients $E_{nm}$ for the smooth residual term $\Ical_{\Rcal}$ used in the high-degeneracy regime $v_{\rm degen} < v < v_{0}$.
	}
	\label{tab:new_R_background_a}
\end{table*}

\begin{table*}[t]
\centering
\scriptsize
\setlength{\tabcolsep}{4pt}
\renewcommand{\arraystretch}{1.12}
\begin{tabular}{rrrrrr}
\hline
$n$ & $m=0$ & $m=1$ & $m=2$ & $m=3$ & $m=4$ \\
\hline
$0$ & $3.3149049275341809$ & $-9.6107276476875878$ & $-2.4836943053595624$ & $0.16521502352544884$ & $0.026740096824277672$ \\
$1$ & $5.4319060481652439$ & $3.8520282412494158$ & $-1.1899042219766041$ & $-0.40240836790386747$ & $0.0090537694271415892$ \\
$2$ & $-0.47361929687725368$ & $2.7760415654510928$ & $0.36584634019033963$ & $0.010693315595553778$ & $-0.0040460192979467763$ \\
$3$ & $-0.4766651801300078$ & $0.090401484039508234$ & $0.66732100884324452$ & $0.038877958799995942$ & $-0.022232883032293153$ \\
$4$ & $-0.19087563446073694$ & $-0.44546229927174202$ & $-0.0087227641722226641$ & $0.10846203781952277$ & $0.0089482745826564951$ \\
\hline
\end{tabular}
\caption{
	Chebyshev coefficients $F_{nm}$ for the regular Kohn region contribution $\Ical_{\Kcal}^{\rm reg}$ used in the high-degeneracy regime $v_{\rm degen} < v < v_{0}$.
}
\label{tab:new_kohn_core_coefficients}
\end{table*}

\end{widetext}